\documentclass[sts]{imsart}

\RequirePackage{amsthm,amsmath,amsfonts,amssymb}
\RequirePackage[authoryear]{natbib}
\RequirePackage{graphicx}

\usepackage[utf8]{inputenc}
\usepackage{enumitem}
\usepackage{verbatim}
\usepackage{array}
\usepackage{bbm}
\usepackage{fancyvrb}
\usepackage{xcolor}
\usepackage{url}
\usepackage{comment}
\usepackage{caption}
\usepackage{subcaption}
\usepackage{float}
\usepackage[colorlinks=true, allcolors=blue]{hyperref}
\usepackage{cleveref}

\usepackage{tkz-graph}  
\usetikzlibrary{shapes.geometric}%

\usepackage{tikz}
\usepackage{graphicx}
\usepackage{mathtools}
\usetikzlibrary{shapes,decorations,arrows,calc,arrows.meta,fit,positioning}
\tikzset{
    -Latex,auto,node distance =1 cm and 1 cm,semithick ,
    state/.style ={ellipse, draw, minimum width = 0.7 cm},
    point/.style = {circle, draw, inner sep=0.04cm,fill,node contents={}},
    bidirected/.style={Latex-Latex,dashed},
    el/.style = {inner sep=2pt, align=left, sloped}
}
\DeclareSymbolFont{AMSb}{U}{msb}{m}{n}
\usetikzlibrary{shapes.arrows,shapes.geometric,
shapes.multipart,decorations.pathmorphing,positioning,
swigs} 

\startlocaldefs
\theoremstyle{plain}
\newtheorem{lem}{Lemma}

\newcommand{\alf}{\alpha}
\newcommand{\bet}{\beta}

\newcommand{\del}{\delta}

\newcommand{\sig}{\sigma}

\newcommand{\kap}{\kappa}

\newcommand{\bfmu}{\boldsymbol{\mu}}
\newcommand{\bfalp}{\boldsymbol{\alpha}}
\newcommand{\bfbet}{\boldsymbol{\beta}}

\newcommand{\bfrho}{\boldsymbol{\rho}}
\newcommand{\bfx}{\boldsymbol{x}}
\newcommand{\bfX}{\boldsymbol{X}}
\newcommand{\bfxi}{\boldsymbol{\xi}}
\newcommand{\bfeta}{\boldsymbol{\eta}}
\newcommand{\bfzeta}{\boldsymbol{\zeta}}
\newcommand{\bfphi}{\boldsymbol{\phi}}
\newcommand{\bfY}{\mathbf{Y}}
\newcommand{\bfZ}{\mathbf{Z}}
\newcommand{\bfW}{\mathbf{W}}
\newcommand{\dd}{\text{d}}
\newcommand{\zk}{\boldsymbol{0}_K}
\newcommand{\bfU}{\mathbf{U}}
\newcommand{\bfV}{\mathbf{V}}
\newcommand{\bfM}{\mathbf{M}}

\newcommand{\bfEpsilon}{\boldsymbol{E}}
\newcommand{\bfEta}{\boldsymbol{H}}

\newcommand{\simiid}{{\,\, \overset{\text{iid}}\sim} \,\,}

\newcommand{\R}{\mathbb{R}}

\newcommand{\F}{\mathcal{F}}

\newcommand{\E}{\mathbb{E}}

\newcommand{\Ber}{\mathsf{Ber}}
\newcommand{\normal}{\mathsf{N}}

\DeclareMathOperator{\tr}{tr}

\newcommand{\mvn}{\mathsf{MVN}}
\newcommand{\bfOmega}{\boldsymbol{\Omega}}

\newcommand{\reals}{\mathbb{R}}
\newcommand{\argmin}{\operatornamewithlimits{argmin}}
\newcommand{\KL}[2]{\text{KL}\left(#1 \mid\mid #2\right)}

\newcommand{\ts}[1]{\textsuperscript{#1}}

\endlocaldefs

\begin{document}

\begin{frontmatter}
\title{On Data Analysis Pipelines and Modular Bayesian Modeling}
\runtitle{On Data Analysis Pipelines and Modular Bayesian Modeling}

\begin{aug}
\author[A]{\fnms{Erin}~\snm{Lipman}\ead[label=e1]{erlipman@uw.edu}}
\and
\author[B]{\fnms{Abel}~\snm{Rodriguez}\ead[label=e2]{abelrod@uw.edu}}

\address[A]{Erin Lipman is PhD Student, Department of Statistics,
University of Washington, Seattle, U.S.\printead[presep={\ }]{e1}.}
\address[B]{Abel Rodriguez is Professor, Department of Statistics,
University of Washington, Seattle, U.S.\printead[presep={\ }]{e2}.}
\end{aug}

\begin{abstract}
The most common approach to implementing data analysis pipelines involves obtaining point estimates from the upstream modules and then treating these as known quantities when working with the downstream ones. This approach is straightforward, but it is likely to underestimate the overall uncertainty associated with any final estimates.  An alternative approach involves estimating parameters from the modules jointly using a Bayesian hierarchical model, which has the advantage of propagating upstream uncertainty into the downstream estimates. However, when modules are misspecified, such a joint model can behave in unexpected ways. Furthermore, hierarchical models require the development of ad-hoc computational implementations that can be laborious and computationally expensive.  Cut inference modifies the posterior distribution to prevent information flow between certain parameters and provides a third alternative for statistical inference in data analysis pipelines. This paper presents a unified framework that encompasses two-step, cut, and joint inference in the context of data analysis pipelines with two modules and uses two examples to illustrate the tradeoffs associated with these approaches.  Our work shows that cut inference provides both some level of robustness and ease of implementation for data analysis pipelines at a lower cost in terms of statistical inference.
\end{abstract}


\end{frontmatter}

\section{Introduction}

Data analysis pipelines, in which a series of algorithms and/or statistical procedures are strung together in such a way that the output generated by those upstream is used as input for those downstream, are widely used across the sciences.  For example, in political science, constructs such as legislators' revealed preferences are often used as either responses or covariates in regression models designed to test hypotheses about legislative behavior (e.g., see \citealp{schickler2000institutional}, \citealp{schwindt2004gender} and \citealp{clark2012examining}).

Two-step (or, more generally, multi-step) procedures, in which the products transferred from one module to the next are limited to point estimators, offer both simplicity and modularity in the implementation of data analysis pipelines.  Simplicity is particularly appealing to practitioners:  a data analysis pipeline can often be put together using existing code and no more than a few function calls.  Modularity implies that it is often easy to update the pipeline or perform sensitivity analysis by simply replacing one or two function calls.  However, an important drawback of multi-step approaches is that they often ignore the uncertainty associated with the output produced by the upstream modules.  For example, the point estimates of legislators' preferences derived from their voting records are typically plugged directly into regression models and treated as ``known'' values even though they are only estimates and, therefore, inherently noisy.  This often leads to under-coverage and bias downstream (e.g., see \citealp{murphy2002estimation} and \citealp{fernandez2011bias}).

An alternative approach to data analysis pipelines is full (Bayesian) hierarchical modeling.  The posterior distribution arising from a hierarchical model that incorporates all of the modules will, by construction, fully account for all the uncertainty associated with the analysis.  One example of such hierarchical models is presented in \cite{lipman2023explaining}, who consider the problem of identifying covariates that might explain differences in legislators' preferences across various types of votes (e.g., procedural vs.\ substantive votes).  Their model can be seen as consisting of two modules, with the first one designed to infer legislators' preferences and whether they differ across vote types, and the second one designed to explain the existence of such a difference as a function of various covariates (please see Section \ref{se:casestudy}).  Yet, working with large hierarchical models can be challenging.  From a computational point of view, combining code that was originally created for the individual modules is more challenging that creating a traditional pipeline.  More importantly, it is well known that misspecification of even a single module in a hierarchical model can cause incorrect estimation of other modules, even if these other modules are correctly specified \citep{plummer2015cuts,liu2022stochastic}, a phenomenon sometimes called \textit{contamination}.

So-called cut posteriors \citep{liu2009modularization,plummer2015cuts,jacob2017better,liu2022general} provide a middle ground between full and multi-step analysis in Bayesian settings.  Cut posteriors modify the posterior distribution so that information is prevented from flowing along certain directions on the graphical model describing the relationships between the parameters and data sources.  Most work on cut posteriors has focused on situations in which there are two separate sets of response variables that provide information about a common set of parameters.  In that kind of setting, one of the two modules / data sources might be considered more reliable, and the cut posterior is sometimes justified as providing a mechanism to prevent the less reliable module from contaminating the other.  

In this paper we investigate the use of cut posteriors in two-module data analysis pipelines. 
This setting is of interest not only because pipelines are widely used in practice, but also because the underlying graphical model has a structure that is different from that of most of the examples in which cut posteriors have been studied before.  To fix ideas, we consider two concrete examples.  In the first one, each of the modules in the pipeline corresponds to a linear matrix-variate regression model.  This setting allows for the derivation of closed-form expressions for the all posteriors, enabling us to clearly illustrate the statistical trade offs associated with various  approaches to data analysis pipelines.  For our second example, we develop a feedback-cut version of the \cite{lipman2023explaining} model and compare its performance against the full model as well as against a two-step procedure.  While no closed-form expressions for the estimates are available in this second example, the non-linearities implicit in the model and the focus on variable selection provide an opportunity to gain additional insight into the behavior all of three approaches.

The remainder of the paper is organized as follows:  Section \ref{se:cuttwomodule} provides a general framework for describing and analyzing data pipelines with two modules. Section \ref{se:matrixvariatelinear} presents our first illustration which focuses on modules that correspond to matrix-variate linear models.  In this context, closed-form expressions are available for the full, two-step and cut distributions, facilitating comparisons. Then, Section~\ref{se:casestudy} introduces a more complex two-module pipeline motivated by an application in political science and compares the performance of all three approaches to inference in its context. Finally, Section~\ref{sec:discussion} summarizes the main takeaway points from the two illustrations and suggests future directions for research.


\section{Data analysis pipelines with two modules}\label{se:cuttwomodule}

Let $\bfY$ represent the response variable. 
We assume that the joint distribution of $\bfY$ can be described by a probability model $p(\bfY \mid \bfphi_1, \bfzeta, \bfW)$ indexed by a set of (known) covariates $\bfW$ and two sets of parameters:  $\bfzeta$, denoting the key parameters of interest in the first module, and $\bfphi_1$, the nuisance parameters.  We further assume that a prior distribution for the nuisance parameters given the parameters of interest and the covariates, $p(\bfphi_1  \mid \bfzeta, \bfW)$, is available.  We refer to the resulting marginal distribution, 
\begin{align}\label{eq:module1}
  p(\bfY \mid \bfzeta, \bfW) = \int p(\bfY \mid \bfphi_1, \bfzeta, \bfW ) p(\bfphi_1 \mid \bfzeta, \bfW) \dd \bfphi_1  ,
\end{align}
as the first module in the pipeline.

\begin{figure}
    \tikzstyle{every edge}=[draw,>=stealth',semithick]

    \centering
        \begin{tikzpicture}[-,>=stealth',auto,
        circularnode/.style={circle,draw,minimum size=8mm,font=\large},
        squarednode/.style={rectangle,draw,minimum size=8mm,font=\large},node distance=1.2cm,scale=1.0,transform shape]
        
            \node[squarednode] (1) at (0,0) {$\bfX$};
            
            \node[circularnode] (5) [above = of 1] {$\bfxi$};
            \node[circularnode] (6) [right = of 5] {$\bfzeta$};
            \node[circularnode] (7) [right = of 6] {$\bfY$};
            \node[squarednode] (4) [below = of 6] {$\mathbf{W}$};
                        
            \path[->] (1) edge (5);
            \path[->] (1) edge (6);
            \path[->] (5) edge (6);
            \path[->] (4) edge (7);
            \path[->] (6) edge (7);
            \path[->] (4) edge (6);
        \end{tikzpicture}
   \caption{Directed acyclic graph representing a generic data analysis pipeline with two modules.}
    \label{fig:dag}
\end{figure}
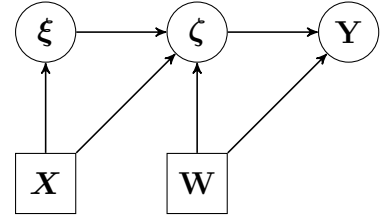

The second module is defined by treating the latent vector $\bfzeta$ as if it were observed.  Again, we assume that the joint distribution for $\bfzeta$ is given by a probability model $p(\bfzeta \mid \bfphi_2, \bfxi, \bfX)$, where $\bfxi$ is the key parameter of interest in the second module, $\bfphi_2$ is a nuisance parameter, and $\bfX$ is a set of auxiliary data (for example covariates used in a regression model). As before, we assume that a prior $p(\bfphi_2 \mid \bfxi, \bfX)$ is available for the nuisance parameters.  The second module of our pipeline is then given by
\begin{align}\label{eq:module2}
  p(\bfzeta \mid \bfxi, \bfX) = \int p(\bfzeta \mid \bfphi_2, \bfxi, \bfX) p(\bfphi_2 \mid \bfxi, \bfX) \dd \bfphi_2  .
\end{align}
Implicit in this definition of the second module is the assumption that the parameters $(\bfxi, \bfphi_2)$ are independent from the data $\bfY$ conditioned on $\bfzeta$.  Figure~\ref{fig:dag} shows the directed acyclic graph (DAG) associated with this two-module model.

The ultimate goal of our analysis is to learn $\bfxi$ from the observed $\bfY$.  Within a Bayesian framework, the most natural approach to answer this question is to base inferences on the marginal posterior distribution,
\begin{equation}\label{eq:fullgen}
    \begin{aligned}
    p_f(\bfxi \mid \bfY) &= \frac{
    \int p(\bfY \mid \bfzeta) p(\bfzeta \mid \bfxi) p(\bfxi) \dd \bfzeta
    }{\int \int p(\bfY \mid \bfzeta) p(\bfzeta \mid \bfxi) p(\bfxi) \dd \bfzeta \dd \bfxi} \\
    &= \int p(\bfxi \mid \bfzeta) p(\bfzeta \mid \bfY) \dd \bfzeta,
    \end{aligned}
\end{equation}
where 
\begin{align*}
p(\bfxi \mid \bfzeta) &= \frac{p(\bfzeta \mid \bfxi) p(\bfxi)}{\int p(\bfzeta \mid \bfxi) p(\bfxi) \dd \bfxi} ,     \\ p(\bfzeta \mid \bfY) &=
\frac{p(\bfY\mid\bfzeta)p(\bfzeta)}{\int p(\bfY\mid\bfzeta)p(\bfzeta) \dd \bfzeta} \\&= \frac{p(\bfY\mid\bfzeta) \int p(\bfzeta \mid \bfxi) p(\bfxi) \dd\bfxi}{\int \int p(\bfY \mid \bfzeta) p(\bfzeta \mid \bfxi) p(\bfxi) \dd \bfxi \dd  \bfzeta}    ,   
\end{align*}
and $p(\bfxi)$ is an appropriate prior on $\bfxi$. Note that, to simplify notation, we have dropped the explicit dependence on $\bfW$ and $\bfX$.  

While this approach is natural from a Bayesian perspective and Markov chain Monte Carlo (MCMC) algorithms to carry out inference under this full model can usually be derived, it can be challenging to implement for practitioners and may be computationally intensive.  This is often true even when using probabilistic programming languages such as \texttt{Stan} \citep{rstan},  \texttt{RBugs} \citep{rbugs} or \texttt{NIMBLE}, \citep{de2017programming}, especially in settings where the parameters of interest are discrete (such as when $\bfxi$ indexes a collection of models).  Furthermore, it has been noted in the literature that, when at least one of the models is misspecified, inferences obtained from the joint model can be nonsensical (e.g., see \citealp{liu2009modularization,plummer2015cuts,jacob2017better,liu2022general}).  We provide an additional example of this in Section \ref{se:casestudy}.

An alternative approach, which is popular among practitioners because of its ease of implementation, involves first  generating a point estimate $\hat{\bfzeta}$ for $\bfzeta$ based on \eqref{eq:module1} (potentially, but not necessarily, through a Bayesian approach to inference), and then treating that point estimator as if it where the true observed value of $\bfzeta$ to then estimate $\bfxi$ from \eqref{eq:module2} (again, possibly through a Bayesian procedure).  More specifically, introduce a working prior $\bar{p}(\bfzeta)$ that is independent of the parameter of interest $\bfxi$. 
We can combine \eqref{eq:module1} with $\bar{p}(\bfzeta)$ to obtain the working first-level posterior  
\begin{align}\label{eq:twostepgen}
\bar{p}(\bfzeta \mid \bfY) = \frac{p(\bfY \mid \bfzeta)\bar{p}(\bfzeta)}{\bar{p}(\bfY)} ,    
\end{align}
where $\bar{p}(\bfY) = \int p(\bfY \mid \bfzeta)\bar{p}(\bfzeta) \dd \bfzeta$.  Given an appropriate loss function $L(\bfzeta, \bfzeta')$, this posterior distribution can be used to generate a point estimator
$$
\hat{\bfzeta}(\bfY) = \arg\min_{\bfzeta'} \int L(\bfzeta, \bfzeta') \bar{p}(\bfzeta \mid \bfY) \dd \bfzeta' .
$$
Then, inferences for $\bfxi$ can be based on the working second-level posterior
\begin{align*}
p_t(\bfxi \mid \bfY) &= p\left(\bfxi \mid \bfzeta
= \hat{\bfzeta}(\bfY) \right) 
\\&= \frac{p\left(\hat{\bfzeta}(\bfY) \mid \bfxi \right) p\left(\bfxi\right)}{\int p\left(\hat{\bfzeta}(\bfY) \mid \bfxi \right) p\left(\bfxi\right) \dd \bfxi} 
\\&= \int p(\bfxi \mid \bfzeta) \delta_{\hat{\bfzeta}(\bfY)}(\bfzeta) \dd \bfzeta ,
\end{align*}
where $\delta_{a}(\cdot)$ denotes the degenerate distribution placing all its mass at $a$.  The joint pseudo-posterior distribution for $(\bfzeta, \bfxi)$ that is implied by this approach is then given by 
\begin{align}\label{eq:pseudojoint_t}
p_t( \bfzeta, \bfxi\mid \bfY) = p\left(\bfxi \mid \bfzeta = \hat{\bfzeta}(\bfY) \right) \bar{p}(\bfzeta \mid \bfY).
\end{align}

Two-step approaches are much easier to implement than the full model approach:  in many settings, software is readily available to carry out inferences based on the first and second level working posteriors, so the whole pipeline can be implemented using just a lines of code.  Additionally, estimating the two models separately may be less computationally intensive than estimating the full, joint model (e.g., when point estimators from the either module can be obtained in closed form, or when the mixing times of the individual MCMC algorithms are much lower than that of the MCMC for the hierarchical model). However, by treating $\hat{\bfzeta}$ as known rather than as the estimate it is, the uncertainty in the estimation of $\bfzeta$ is not propagated through to the second module. This leads to underestimation of the uncertainty in the estimation of $\bfxi$ and, potentially, to bias as well. Furthermore, the choice of the point estimate $\hat{\bfzeta}$ based on the first-level posterior $\bar{p}(\bfzeta \mid \bfY)$ is dependent on the choice of loss function.  This is particularly problematic because it is not always clear how to select the most appropriate loss function, especially since the problem of estimating $\bfzeta$ is not necessarily of primary interest.  For example, when $\bfzeta$ is a vector of binary indicators, it is common to use a zero-one loss, resulting in point estimates that threshold the posterior probabilities, with the threshold depending on the relative loss assigned to a false positive versus a false negative. The point estimate $\hat{\bfzeta}$ (and, therefore, the results from the overall procedure) can be highly dependent on the exact value chosen for these relative losses.  While a threshold of 0.5 (which implies equal losses for both types of errors) is a common default choice, it is not clear that it is appropriate in every setting.

A third  approach is to consider a cut posterior. Its definition starts with the same working first-level posterior $\bar{p}(\bfzeta \mid \bfY)$ as the two-step approach (recall Equation \eqref{eq:twostepgen}), which is used to construct the joint pseudo-posterior
\begin{align}\label{eq:pseudojoint_c}
p_c(\bfzeta, \bfxi \mid \bfY) = p(\bfxi \mid \bfzeta) \bar{p}(\bfzeta \mid \bfY).     
\end{align}
The marginal cut posterior for $\bfxi$ is then given by
\begin{align}\label{eq:cutgen}
    p_c(\bfxi \mid \bfY) = \int p(\bfxi \mid \bfzeta) \bar{p}(\bfzeta \mid \bfY) \dd \bfzeta  .
\end{align}

While cut posteriors in general require more sophisticated computational approaches than two-step posteriors (e.g., see \citealp{plummer2015cuts}), the case of data analysis pipelines admits a straightforward computational approach that reduces to first obtaining $B$ samples 
$$\bfzeta^{(1)},\dots,\bfzeta^{(B)}$$ 
from the first level posterior, and then obtaining samples from the second stage posterior by generating $\bfxi^{(b)}\sim p(\bfxi\mid\bfzeta^{(b)})$.
In practice, this involves running a separate MCMC algorithm for each posterior sample $\bfzeta^{(b)}$ for a relatively short stretch of time.  While doing so is more computationally intensive than running a two-step procedure, it is often as easy to implement using pre-existing software.  Furthermore, we note that such a procedure can be naively parallelized since the samples $\bfxi^{(b)}$ are independent given the samples from the first-stage posterior. 

The previous discussion makes it clear that $p_f(\bfxi \mid \bfY)$, $p_c(\bfxi \mid \bfY)$ and $p_t(\bfxi \mid \bfY)$ can all be written as mixtures of posterior distribution for the second module, $p(\bfxi \mid \bfzeta)$, with respect to slightly different mixing distributions (the ``coherent'' posterior $p(\bfzeta \mid \bfY)$, the incoherent first-level working posterior $\bar{p}(\bfzeta \mid \bfY)$, and a point mass at a point estimator obtained from the first-level working posterior, respectively).  Furthermore, because both $p_t$ and $p_c$ are based on the working posterior, we do not have to worry about conflicts between the modules as they relate to $\bfzeta$.  
However, unlike $p_t$, $p_c$ does allow us to propagate (some) uncertainty from the estimation of $\bfzeta$ to the second stage. 

\subsection{The cut posterior as an optimal approximation to the full posterior under constraints}\label{se:approxcut}
Previously, we suggested that cut posteriors might be better than two-step approaches in terms of capturing the true uncertainty associated with the model parameters.  In this section, we provide a formal justification for this statement.  To do so, consider the class of distributions
$$
\F = \{q(\bfzeta,\bfxi) : 
\int q(\bfzeta,\bfxi)d\bfxi = \bar{p}(\bfzeta \mid \bfY)\},
$$
i.e., the class of all probability distributions for $(\bfzeta, \bfxi)$ that have the first-level working posterior as their marginal for $\bfzeta$.  The following theorem, which is adapted from \cite{yu2023variational}, shows that $p_c(\bfzeta,\bfxi)$ is the best approximation to the full posterior $p_f(\bfzeta,\bfxi)$ with the class $\F$ (in the Kullback–Leibler sense):
\begin{lem}\label{lem:KL}
    For the cut posterior
    \begin{enumerate}
        \item $p_c(\bfzeta,\bfxi\mid \bfY)
        =\argmin\limits_{q\in \F}
        \KL{q(\bfzeta,\bfxi)}{p_f(\bfzeta,\bfxi\mid \bfY)}$
        \item 
        $\KL{p_c(\bfzeta,\bfxi\mid \bfY)}{p_f(\bfzeta,\bfxi\mid \bfY)}
        \\ \qquad=\KL{\bar{p}(\bfzeta \mid \bfY)}{p_f(\bfzeta\mid \bfY)}$
    \end{enumerate}
\end{lem}

The proof can be seen in Section 1 of the supplementary materials 
Because both $p_t(\bfzeta,\bfxi)$ and $p_c(\bfzeta,\bfxi)$ both belong to $\F$, this result suggests that, at least in this sense, cut posteriors are superior to two-step posteriors.

\subsection{Relationship with other relevant literature}\label{se:litcomparison}

The graphical model in Figure \ref{fig:dag} is very similar to the graphical model that appears in the context of many selective inference problems \citep{kuchibhotla2022post,taylor2015statistical,berk2013valid,benjamini2005false,hjort2003frequentist}.  Selective inference concerns itself with designing procedures with valid statistical properties when carried out after model / hyperparameter (e.g., LASSO penalty) selection.  As this description suggests, the focus of selective inference procedures is on how uncertainty downstream the pipeline (on $\bfxi$) affects inferences carried out upstream the pipeline (on $\bfzeta$).  Our focus in this paper is the opposite: we are interested in the effect that uncertainty upstream in the pipeline (on $\bfzeta$) affects inferences downstream (on $\bfxi$).  Frequentist work on this version of the problem is, as far as we are aware, limited.

\begin{figure}
    \centering
        \tikzstyle{every edge}=[draw,>=stealth',semithick]
        \begin{center}
        \begin{tikzpicture}[-,>=stealth',auto,
        circularnode/.style={circle,draw,minimum size=8mm,font=\large},
        squarednode/.style={rectangle,draw,minimum size=8mm,font=\large},node distance=1.2cm,scale=1.0,transform shape]
            \node[circularnode] (1) at (0,0) {$\bfY$};
            \node[circularnode] (2) [left = of 1] {$\bfZ$};
            \node[circularnode] (3) [below = of 1] {$\bfzeta$};
            \node[circularnode] (4) [below = of 2] {$\bfxi$};
            \node[squarednode] (5) [right = of 3] {$\bfW$};
            \node[squarednode] (6) [left = of 4] {$\bfX$};
            
            \path[->] (3) edge (1);
            \path[->] (4) edge (1);
            \path[->] (4) edge (2);
            \path[->] (5) edge (1);
            \path[->] (5) edge (3);
            \path[->] (6) edge (2);
            \path[->] (6) edge (4);
            
        \end{tikzpicture}
        \end{center}        
    \caption{DAG commonly used in literature on feedback cut. This contrasts with the linear DAG from our pipeline setting.}
    \label{fig:dag2}
\end{figure}
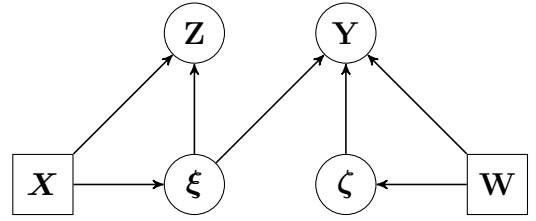

On the other hand, most applications of cut posteriors that we are aware of focus on models that can be described with a DAG similar to that presented in Figure \ref{fig:dag2}.  A key difference is the presence of $\bfZ$, which corresponds to another set of response variables that provide information about $\bfxi$.  A second difference is the structure of the relationship between $\bfxi$ and $\bfzeta$. These differences is the structure of the underlying graph have  implications on how the cut posterior is defined and justified, and they make our problem distinct from those that have been discussed in the literature previously.

\section{Illustration 1:  Matrix-variate Linear Models}\label{se:matrixvariatelinear}

We consider first a version of the setup in Section \ref{se:cuttwomodule} in which $\bfY$ corresponds to an $N \times J$ matrix of continuous responses and the first-level module corresponds to a matrix-variate linear regression
\begin{align*}
\bfY = \bfzeta \bfW + \bfEpsilon ,
\end{align*}
where $\bfW$ is a $L \times J$ full-rank design matrix, $\bfzeta$ is a $N \times L$ matrix of unknown regression coefficients ($L \ll J$), and $\bfEpsilon$ is an $N \times J$ matrix of observational errors, which are assumed to follow a matrix-variate normal distribution, $\bfEpsilon \sim \mvn_{N,J} ( \mathbf{0}, \mathbf{I}_{N}, \sigma^2 \mathbf{I}_{J})$ (please see Section 2 of the supplementary materials). 
Note that we use  $\mathbf{I}_{n}$ to denote the $n \times n$ identify matrix.  For the second module, we consider again a matrix-variate linear regression,
\begin{align*}
\bfzeta = \bfX \bfxi + \bfEta ,
\end{align*}
where $\bfX$ is an $N \times K$ full-rank design matrix, $\bfxi$ is a $K \times L$ matrix of regression coefficients ($K \ll N$), and $\bfEta \sim \mvn_{N,L}\left( \mathbf{0},  \tau^2 \mathbf{I}_{N}, \mathbf{I}_{L}\right)$.

In the sequel, we assume that both $\sigma^2$ and $\tau^2$ are known.  This implies that there are no nuisance parameters $\bfphi_1$ or $\bfphi_2$. 
Furthermore, we consider the reference prior for $\bfxi$, $p(\bfxi) \propto 1$, as well as for the working prior $\bfzeta$, $\bar{p}(\bfzeta) \propto 1$.  In this setting, the full, two-step and cut posteriors can be computed in closed form.  We start by discussing the full posterior.  Note that, in this setting, we can explicitly integrate $\bfzeta$,  resulting in 
$\bfY \mid \bfxi \sim \mvn_{N,J} \left( \bfX \bfxi \bfW, \mathbf{I}_{N}, \bfOmega\right)$, where $\bfOmega = \sigma^2 \mathbf{I}_J  + \tau^2 \bfW^T  \bfW$.  Hence, under the reference prior (see Section 2.1 of the supplementary materials), 

\begin{align}\label{eq:linear_fullpost}
p_f\left(\bfxi \mid \bfY  \right) = \mvn_{K,L} \left( \bfM_f, \bfU_f, \bfV_f \right), 
\end{align}
where
\begin{align*}
    \bfM_f &= [ \bfX^T \bfX ]^{-1} \bfX^T \bfY \bfOmega^{-1} \bfW^T \left[ \bfW \bfOmega^{-1} \bfW^T \right]^{-1}, \\
    \bfU_f &= [\bfX^{T} \bfX]^{-1}, \\
    \bfV_f &= \sigma^2 [\bfW \bfW^T]^{-1} + \tau^2 \mathbf{I}_{L}.
\end{align*}

Consider now the two-step and cut posteriors.  Under the working prior $p(\bfzeta) \propto 1$, the first-level working posterior is 
$$\bfzeta \mid \bfY \sim \mvn_{N,L} \left( \bfY \bfW^T [ \bfW \bfW^T ]^{-1}, \mathbf{I}_{N}, \sigma^2 [ \bfW \bfW^T ]^{-1} \right).$$  
A similar calculation leads to 
\begin{align*}
\bfxi \mid \bfzeta \sim \mvn_{K,L} \left( [\bfX^T \bfX]^{-1}\bfX^T\bfZ ,  [\bfX^T\bfX]^{-1}, \tau^2 \mathbf{I}_{L} \right)
\end{align*}
(see Section 2.2 of the supplementary materials).

To construct the two-step posterior, a natural choice for a point estimator is the posterior mean, 
$$
\hat{\bfzeta}(\bfY) = \bfY \bfW^T [ \bfW \bfW^T ]^{-1},
$$
which is optimal under squared error loss.  This results in
\begin{align}\label{eq:linear_twosteppost}
p_t(\bfxi \mid \bfY) = \mvn_{K,L} \left(\bfM_t,\bfU_t,\bfV_t\right) .
\end{align}
where
\begin{align*}
    \bfM_t &= [\bfX^T \bfX]^{-1}\bfX^T\bfY \bfW^T [ \bfW \bfW^T ]^{-1}, \\
    \bfU_t &= [\bfX^T\bfX]^{-1}, \\
    \bfV_t &= \tau^2 \mathbf{I}_{L}.
\end{align*}

In contrast, the cut posterior is obtained by integrating over the first-level working posterior (see Appendix \ref{ap:derivationcutpost}), resulting in 
\begin{align}\label{eq:linear_cutpost}
p_c\left(\bfxi \mid \bfY\right) = \mvn_{K,L} \left(  \bfM_c,\bfU_c,\bfV_c\right),
\end{align}
where
\begin{align*}
    \bfM_c &= [\bfX^T \bfX]^{-1}\bfX^T\bfY \bfW^T [ \bfW \bfW^T ]^{-1}, \\
    \bfU_c &= [\bfX^T \bfX]^{-1}, \\
    \bfV_c &= \sigma^2 [\bfW \bfW^T]^{-1} + \tau^2 \mathbf{I}_{L}.    
\end{align*}

The relative simplicity of these expressions allows us to understand the trade-offs associated with these various posterior distributions.  First, note that the posterior mean for $\bfxi$ is the same under $p_t$ and $p_c$, yielding (under squared error loss) the point estimator 
$$\tilde{\bfxi}(\bfY) = [\bfX^T \bfX]^{-1}\bfX^T\bfY \bfW^T [ \bfW \bfW^T ]^{-1}.$$
This is an unbiased estimator of $\bfxi$.  When the variances $\tau^2$ and $\sigma^2$ are correctly specified, this estimator has a higher variance than the estimator that arises from $p_f$, 
$$\hat{\bfxi} (\bfY) =  [ \bfX^T \bfX ]^{-1} \bfX^T \bfY  \bfOmega^{-1} \bfW^T \left[ \bfW \bfOmega^{-1} \bfW^T \right]^{-1}$$ 
(which is the best linear unbiased estimator in this setting).  On the other hand, when the values of $\tau^2$ and $\sigma^2$ are misspecified (i.e., when the values used to fit the model are different from those in the underlying true data generation mechanism), both $\hat{\bfxi}(\bfY)$, $\tilde{\bfxi}(\bfY)$ remain unbiased but the fact that $\tilde{\bfxi}(\bfY)$ is independent of $\bfOmega$ means that the point estimator is robutst to this kind of model misspecification.  In contrast, the variance of $\hat{\bfxi}(\bfY)$ can potentially be very large (and, in particular, larger than that of $\tilde{\bfxi}(\bfY)$) in cases where the ratio $\sigma^2/\tau^2$ used to fit the model is very different from the true ratio (see Sections 2.1 and 2.2 of the supplementary materials).
This contrasting behavior illustrates how the use of cut posteriors can provide robustness to some types of model misspecification, which might reside on any (or both) of the modules.

Next, note that the posterior variances for $p_f$ and $p_c$ are the same, and that they are larger than the variance of $p_t$, in the sense that 
\begin{align*}
\tr \left( \left[\sigma^2 [\bfW \bfW^T]^{-1} + \tau^2 \mathbf{I}_{L} \right] \otimes [\bfX^T \bfX]^{-1} \right) 
\\\quad> \tr \left( \tau^2 \mathbf{I}_{L} \otimes [\bfX^T \bfX]^{-1} \right) .
\end{align*}
This shows how one might expect two-step posteriors to underestimate the uncertainty associated with the estimates of $\bfxi$.  Together, these two observations also suggest that cut posteriors offer the best trade offs between reasonably accurate uncertainty estimation, efficiency in point estimation, and robustness to certain types of model misspecification.

Finally, note that, when $\sigma^2/\tau^2 \to 0$ (i.e., when the observational noise in the second module dominates), all three posteriors are very close to each other and, in the limit, are identical to $p_t$.  This suggests that there are situations in which it might not matter much which procedure is used, but that those situations might require very stringent conditions on the data generation mechanism that should be checked for any particular dataset.



\section{Illustration 2:  Hypothesis testing in hierarchical models with latent variables}\label{se:casestudy}

The previous illustration is useful in providing some intuition about the behavior of the various types of posterior distributions we discussed in Section \ref{se:cuttwomodule}.  However, its relatively simplicity limits the insights that can be gained from it.  In this Section, we consider a more elaborate example that is motivated by our own applied work but where closed-form expressions for the posteriors are not available.  We start by describing the applied problem that motivates the model.

Recovering the preferences of members of deliberative bodies (such as legislatures or judicial panels) from their voting records is an essential task in various areas of application, but specially in political science.  Accordingly, a broad and growing literature on scaling models has developed since the early 1980s (e.g., \citealp{poole1985spatial,poole2005spatial, jackman2001multidimensional,MartinQuinnDynamicIdealPoint2002a,clinton2004statistical,yu2021spatial,duck2022ends,lei2023novel}).  While sometimes used for descriptive purposes (e.g., see \citealp{jenkins2006impact}, \citealp{poole2006ideology}, \citealp{shor2011ideological} and \citealp{luque2022bayesian}), the estimates generated by these models are most often used as inputs for other models whose aim is to test specific substantive theories of member's behavior (e.g., see  \citealp{schickler2000institutional}, \citealp{schwindt2004gender} and \citealp{clark2012examining}).

In this line of work, one recurring question is whether members' preferences can be considered constant across across time or voting domains, and if not, what factors drive the divergent behavior (e.g., see \citealp{MartinQuinnDynamicIdealPoint2002a}, \citealp{jessee2014two}, \citealp{Lof16}, \citealp{moser2021multiple}, \citealp{lei2023dynamic} and \citealp{lipman2023explaining}).
In this paper we focus on the problem of explaining differences in legislator behavior on procedural versus substantive  votes.  Scholars have long been interested in understanding what factors drive legislators to vote differently across domains.  Generally speaking, legislators must balance the competing interests of their constituents (e.g., see \citealp{lope07-strategic,flei04-shrinking,bond02-disappearance}), their political parties (e.g., see \citealp{hix16-governmentopposition}), interest groups (e.g., see \citealp{kau13-congressman}), and their own beliefs.  In the specific case of procedural vs.\ substantive votes, there is broad evidence that factors like party affiliation \citep{patty2010dilatory,carson2014procedural,jessee2014two}, cross-pressure from constituents and party leaders \citep{kirkland2014partisanship}, as well as other forms of electoral considerations \citep{shin2017choice} affect votes on each domain differently.

The data analysis problem at hand can be construed as requiring a two-module pipeline in which the first module is used to identify legislators whose preferences remain constant across domains (which we call ``bridges''), and the second module is used to identify variables that might explain whether a particular legislator is a bridge. \cite{lipman2023explaining} developed a hierarchical formulation to address this question using what we refer to here as the full posterior.  We proceed now to review that model, develop the corresponding two-step and cut posteriors, and explore the differences when all three approaches are applied to various datasets.
 
\subsection{Notation}\label{se:notation}

Let $i=1,\dots,N$ index the set of legislators and $j=1,\dots,J$ index the votes taken during that session. The response variable $\bfY$ is an $N\times J$ binary matrix of roll call voting data, where $y_{i,j}=1$ represents a ``yea'' vote for the $i$\ts{th} legislator on the $j$\ts{th} measure and $y_{i,j}=0$ represents a ``nay'' vote.  Furthermore, let $w_j\in\{0,1\}$ be the a (known) indicator for whether the bill is a final passage vote ($w_j=1$) or a procedural vote ($w_j=0$).

\subsection{The first module}\label{se:module1}

The first module in our pipeline is a very slight variation of the hierarchical model introduced in \cite{Lof16}.  The likelihood for this model is given by
\begin{multline}\label{eq:likelihood2}
y_{i,j} \mid \mu_j, \alpha_j, \beta_{i,0}, \beta_{i,1}, w_j \\
\sim \Ber \left(y_{i,j} \, \bigg | \, \frac{\exp\left\{ \mu_j+\alpha_j\beta_{i,w_j}\right\}}{1 + \exp \left\{ \mu_j+\alpha_j^T\beta_{i,w_j} \right\}} \right),  
\end{multline}
for $i=1,\ldots, N$ and  $j=1,\ldots, J$,
where the parameters $\bet_{i,0},\bet_{i,1} \in \reals$ are the \textit{ideal points} for legislator $i$ on procedural and final passage votes, respectively, and the parameters $\mu_j \in \reals$ and $\alpha_j \in \reals$ are unknown parameters that can be understood as the baseline probability of an affirmative vote and the discrimination associated with measure $j$, respectively.  The ideal points can in turn be interpreted as the preferred outcomes of legislators in the underlying latent policy space. Understanding the differences in the preferences across domains is at the heart of the our application.

The next level of the hierarchy is given by a joint prior on $(\beta_{i,0}, \beta_{i,1})$ that explicitly accounts for the possibility that $\beta_{i,0} = \beta_{i,1}$. Such a prior is defined conditionally on the vector of bridging indicators $\bfzeta = (\zeta_1, \ldots, \zeta_N)$, with $\zeta_i \in\{0,1\}$, so that 
\begin{multline}\label{eq:priorbeta}
  (\beta_{i,0},\beta_{i,1}) \mid \zeta_i, \bfrho_{\beta}, \sig^2_{\beta} 
  \\\simiid 
  \begin{cases}
     \normal\left(\beta_{i,0} \mid \bfrho_{\beta}, \sig^2_{\beta} \right) \, \delta_{\beta_{i,0}}\left(\bfbet_{i,1}\right) & \zeta_i=1 , \\
     \normal\left(\beta_{i,0} \mid\bfrho_{\beta}, \sig^2_{\beta}\right)\normal\left(\beta_{i,1} \mid \bfrho_{\beta}, \sig^2_{\beta}\right) & \zeta_i=0 .\\
\end{cases}
\end{multline}
Note that $\zeta_i = 1$ corresponds to the hypothesis $\beta_{i,0} = \beta_{i,1}$, while $\zeta_i = 0$ corresponds to $\beta_{i,0} \ne \beta_{i,1}$.

The specification of the first module is completed by eliciting priors for the parameters $\bfmu = (\mu_1, \ldots, \mu_J)$ and $\bfalp = (\alpha_1, \ldots, \alpha_J)$ and the hyperparameters $\rho_{\beta}$ and $\sigma_{\beta}^2$ (please see Appendix~\ref{ap:model_description1} for details on these). Referring back to the notation from Section \ref{se:cuttwomodule}, we have $\bfphi_1 = (\bfalp, \bfmu, \bfbet_0, \bfbet_1, \rho_{\beta}, \sigma^2_{\beta})$, and 

\begin{equation}\label{eq:module1_votes}
\begin{aligned}
p(\bfY \mid \bfzeta) &= \\
& \int \left\{ \prod_{i=1}^{N}\prod_{j=1}^{J}   p(y_{i,j} \mid \mu_j, \alpha_j, \beta_{i,0}, \beta_{i,1}, w_j) \right\} 
\\
&\times\left\{ \prod_{i=1}^{N} p(\beta_{i,0},\beta_{i,1} \mid \zeta_i, \rho_{\beta}, \sigma^2_{\beta}) \right\} 
\\  
&\times p(\bfmu)  p(\bfalp) p(\rho_{\beta}) p(\sigma^2_{\beta}) \, \dd \bfmu \, \dd \bfalp \, \dd \bfbet_0 \, \dd \bfbet_1 \, \dd \rho_{\beta} \, \dd \sigma^2_{\beta} .
\end{aligned}
\end{equation}

\subsection{The second module}\label{se:module2}

The second module of the pipeline relates the vector of bridging indicators $\bfzeta$ to various covariates of interest through a logistic regression:
\begin{align}\label{eq:zetalik}
\zeta_i \mid \bfx_i, \eta_0, \bfeta \sim \Ber\left( \zeta_i \, \bigg | \, \frac{1}{1 + \exp\left\{ - \left( \eta_0 + 
\bfx_i^T \bfeta \right)\right\}} \right) ,    
\end{align}
where $\eta_0$ is an intercept, $\bfeta = (\eta_1, \ldots, \eta_K)^T$ is a $K$-dimensional vector of unknown regression coefficients, and $\bfx_i$ is the $i$-th row of the $N\times K$ design matrix $\bfX$.

Similarly to the first module, the prior on $\bfeta$ is defined hierarchically.  Since our ultimate goal is variable selection, let $\bfxi = (\xi_1, \ldots, \xi_K)$ correspond to a vector of indicator variables such that $\xi_k = 1$ if and only if the $k$-th variable of interest is relevant to explaining whether legislators act as bridges.  Then, conditional on $\bfxi$,  
the prior for $\bfeta$ takes the form 
\begin{align*}
p(\bfeta \mid \bfxi) =  N\left(\eta_{\bfxi} \, \Big | \, 0, 4 N \left(\bfX_{\bfxi}^T \bfX_{\bfxi} \right)^{-1}\right)  \prod_{\{ k : \xi_k = 0\}} \delta_{0} (\eta_k) ,
\end{align*}
where $\eta_{\bfxi}$ denotes the subvector of $\bfeta$ that only includes the coefficients for which $\xi_k = 1$, and $\bfX_{\bfxi}$ denotes the matrix whose columns correspond to the variable for which $\xi_k = 1$.  This corresponds to an (approximately) unit information prior for $\bfeta_{\bfxi}$ \citep{sabanes2011hyper}.  To complete the model, the intercept $\eta_0$ is assigned a standard logistic prior (see also next Section).  Then, in the notation of Section~\ref{se:cuttwomodule}, $\bfphi_2 = (\eta_0,\bfeta)$, and the second module is given by
\begin{align}\label{eq:module2_votes}
  p(\bfzeta \mid \bfxi) = \int \prod_{i=1}^{N} p(\zeta_i \mid \eta_0,\bfeta,\bfx_i) p(\bfeta \mid\bfxi,\bfX) p(\eta_0) \dd \bfeta \, \dd\eta_0 .
\end{align}

\subsection{Priors}\label{se:priors}

In this Section, we discuss the prior on the key parameter of interest $\bfxi$, as well as the working prior on $\bfzeta$ to be used in the cut and two-step posteriors.  For $\bfxi$, we employ a Beta-Bernoulli prior such that

\begin{equation}\label{eq:xiprior}
\begin{aligned}
  p(\bfxi) &= \frac{\Gamma\left(\bfxi_{\cdot} + 1 \right)\Gamma\left(K - \bfxi_{\cdot} + 1\right)}{\Gamma( K + 2)} \\
  &= \int_0^1 \upsilon^{\bfxi_{\cdot}} (1 - \upsilon)^{K - \bfxi_{\cdot}} d \upsilon,
\end{aligned}
\end{equation}
where $\Gamma(\cdot)$ denotes the Gamma function and $\bfxi_{\cdot} = \sum_{k=1}^{K} \xi_k$ is the number of predictors included in the model \citep{scott2006exploration,scott2010bayes}.  Similarly, for the working prior $\bar{p}(\bfzeta)$, we also consider a Beta-Binomial distribution
\begin{equation}\label{eq:zetaworkingprior}
  p(\bfzeta) = \frac{\Gamma\left(\bfzeta_{\cdot} + 1 \right)\Gamma\left(N - \bfzeta_{\cdot} + 1\right)}{\Gamma( N + 2)},
\end{equation}
where, as before, $\bfzeta_{\cdot} = \sum_{i=1}^{N} \zeta_i$.  Note that, because we assigned $\eta_0$ a logistic prior, we have
$$
\bar{p}(\bfzeta) = p(\bfzeta \mid \bfxi = (0,0, \ldots, 0)),
$$
i.e., the working prior corresponds to the ``correct'' hierarchical prior when no covariates affect the bridging probability.  In this case, Lemma~\ref{lem:KL} implies that

\begin{equation}\label{eq:klbound}
\begin{aligned}
    & \KL{p_c(\bfzeta,\bfxi\mid \bfY)} 
    {p_f(\bfzeta,\bfxi\mid \bfY)}  \\
    & \;\;\;\;\;\;\;\;\;\;\;\;\;\;\;\;\;\;\;\;\; = \KL{\bar{p}(\bfzeta \mid \bfY)}{p_f(\bfzeta\mid \bfY)}\\
    & \;\;\;\;\;\;\;\;\;\;\;\;\;\;\;\;\;\;\;\;\; = \KL{p_f(\bfzeta \mid \bfY, \bfxi=\zk)}{p_f(\bfzeta\mid \bfY)} \\
    & \;\;\;\;\;\;\;\;\;\;\;\;\;\;\;\;\;\;\;\;\; \le -\log\left(p_f(\bfxi=\zk\mid\bfY)\right) ,
\end{aligned}
\end{equation}
where $0_K$ is the zero vector in $\R^K$.

Hence, $\KL{p_c(\bfzeta,\bfxi\mid \bfY)}{p_f(\bfzeta,\bfxi\mid \bfY)} \to 0$ as $p(\bfxi=\zk\mid\bfY) \to 1$.  This relationship implies a sort of consistency that, we believe, is highly desirable and should be kept in mind when selecting the working prior:  if the full model strongly supports the exclusion of all covariates from the second module, then posterior inferences for the cut posterior will be (nearly) identical to those from obtained from the joint model.

\subsection{Computation}

Posterior inference for all three versions of the posterior distribution relies on MCMC algorithms to produce samples $\left(\bfzeta^{(1)},\bfxi^{(1)}\right), \ldots, \left(\bfzeta^{(B)},\bfxi^{(B)}\right)$ from the posterior distributions. This Section provides a high-level overview of such algorithms that highlights the similarities and differences.  An implementation of the algorithms is available at \url{https://github.com/e-lipman/ModularBayesUSHouse}.

\subsubsection{Full posterior}\label{se:mcmc_full}

Computation for the full posterior uses the MCMC sampler introduced in \citet{lipman2023explaining}. It is useful to view this algorithm as a two-block sampler that alternately samples the parameters from the first module and those from the second module from their respective full conditional distributions. In particular in iteration $b$, we first sample the parameters of the first module from 
\begin{multline}\label{eq:gibbs1}
p\left(\bfzeta^{(b)},\bfphi_1^{(b)} \mid \bfxi^{(b-1)},\bfphi_2^{(b-1)}, \cdots \right) 
\\ \propto p\left( \bfY \mid \bfphi_1^{(b)}, \bfzeta^{(b)}, \right)  p\left(\bfphi_1^{(b)} \mid \bfzeta^{(b)} \right) \\
\times p\left(\bfzeta^{(b)} \mid \bfxi^{(b-1)},\bfphi_2^{(b-1)} \right)
\end{multline}
and then update the module 2 parameters
\begin{multline}\label{eq:gibbs2}
p\left(\bfxi^{(b)},\bfphi_2^{(b)} \mid\bfzeta^{(b)}, \cdots \right) \\
\propto p\left(\bfzeta^{(b)} \mid \bfxi^{(b)},\bfphi_2^{(b)} \right) p\left( \bfphi_2^{(b)} \mid \bfxi^{(b)} \right) p\left( \bfxi^{(b)} \right)
\end{multline}

Sampling from the conditional distribution in \eqref{eq:gibbs1} is done using a series of Gibb's steps that rely on the Polya-Gamma data augmentation approach of \cite{polson2013bayesian}. Sampling from the conditional distribution in \eqref{eq:gibbs2} is achieved using a Metropolis-within-Gibbs sampler 
that again relies on a Polya-Gamma augmentation.

\subsubsection{Two-step posterior}\label{se:mcmc_twostep}

Computation for the two-step posterior relies on slight variations of the same two blockwise samplers described above.  First we generate samples $\bfzeta^{(1)}, \bfzeta^{(2)}, \ldots$ from the first-level working posterior, which in our case reduces to sampling from \begin{multline}\label{eq:gibbs3}
    \bar{p} \left(\bfzeta^{(b)},\bfphi_1^{(b)} \mid \cdots \right) \\
    \propto p\left( \bfY \mid \bfphi_1^{(b)}, \bfzeta^{(b)}, \right)  p\left(\bfphi_1^{(b)} \mid \bfzeta^{(b)} \right) \bar{p}\left(\bfzeta^{(b)} \right)
\end{multline}
The MCMC sampler for \eqref{eq:gibbs3} is extremely similar (but not identical) to that for \eqref{eq:gibbs1} (e.g., see \citealp{Lof16}). We then use a simple zero-one loss function 
\begin{multline*}
    L(\bfzeta, \bfzeta')
    = \\
    \sum_{i=1}^N 
    \left\{a_1 I(\zeta_i=1,\zeta_i'=0) +
    a_2 I(\zeta_i=0,\zeta_i'=1)\right\}
\end{multline*}
to generate our point estimator $\hat{\bfzeta}$, which results in $\hat{\zeta}_i = I\left\{\bar{p}\left(\zeta_i=1 \mid \bfY \right)>\frac{a_2}{a_1+a_2}\right\}$.  As we mentioned earlier, a common choice is $a_1=a_2$, leading to ta threshold of $0.5$ for the posterior probability.  
Finally, samples from $p_t(\bfxi \mid \bfY)$ can be obtained using the same algorithm to sample from \eqref{eq:gibbs2} by simply fixing $\bfzeta^{(b)} = \hat\bfzeta$ for all $b$. 

\subsubsection{Cut posterior}\label{se:mcmc_cut}

Computation for the cut posterior relies on exactly the same algorithms used for the two-step posterior, but their output is used is a slightly different way.  As for the two-step posterior, our first step is to generate samples $\bfzeta^{(1)}, \bfzeta^{(2)}, \ldots$ from \eqref{eq:gibbs3}.  Then, for each of these samples, we run an independent MCMC algorithm to generate pairs 
$$\left(\tilde{\bfxi}^{(b,1)}, \tilde{\bfphi}_2^{(b,1)}\right), \ldots \left(\tilde{\bfxi}^{(b,S)}, \tilde{\bfphi}_2^{(b,S)}\right)$$ from 
\begin{multline*}
p\left(\bfxi^{(b,s)},\bfphi_2^{(b,s)} \mid\bfzeta^{(b)}, \cdots \right) \propto
\\ 
p\left(\bfzeta^{(b)} \mid \bfxi^{(b,s)},\bfphi_2^{(b,s)} \right)
p\left(\bfphi_2^{(b,s)} \mid \bfxi_2^{(b,s)} \right)
p\left(\bfxi^{(b,s)} \right)
\end{multline*}
(recall Equation \eqref{eq:gibbs2}).  For a sufficiently large value of $S$, we then set $\left(\bfxi^{(b)}, \bfphi_2^{(b)}\right) = \left(\tilde{\bfxi}^{(b,S)}, \tilde{\bfphi}_2^{(b,S)}\right)$.

\subsection{Empirical results}

We evaluate the three posterior distribution on the the same data presented in \cite{lipman2023explaining}. 
It consists of roll call votes in the U.S.\ House of Representatives between 1973 and 2014.  We consider 21 explanatory variables for the tendency of individual legislators to reveal different preferences across procedural and final passage votes. In line with the discussion in Section \ref{se:casestudy}, these include constituency-level, legislators-level, and chamber-level covariates (see Appendix~\ref{ap:model_description2}). 

For each of our models, we ran multiple parallel chains in order to assess mixing and convergence using the convergence diagnostic described in \cite{GeRu92}. To sample from the full posterior for each House, we run four chains each with $20,000$ burn-in iterations followed by $300,000$ samples from which we store every $20$ samples for a total of $B=15,000$.\footnote{The posterior distribution for the 105\ts{th} congress shows multimodaility in the size of the selected model. To ensure adequate sampling from each mode, we run 8 chains with 500,000 iterations post burn-in, of which we store every 25 samples for a total of $B=20,000$.}  For the two-step and cut posteriors, we first sample from the working posterior and obtain two chains of $B=5,000$ samples each by first burning $10,000$ iterations and then generating another $50,000$ samples, which are thinned by a factor of 10.  For the two-step posterior, the two chains are combined into a single sample, which is used to generate the point estimator $\hat{\bfzeta}$.  Then, we use the MCMC sampler for the second module to generate two more chains, again by first burning $10,000$ iterations and then generating another $50,000$ samples, which are thinned by a factor of 10.  On the other hand, for the cut posterior, the MCMC sampler for the second module is run for $S=200$ iterations for each sample from the two chains from the working posterior.  A sensitivity analysis using larger values of $S$ shows essentially identical results.  In terms of execution time, it is worthwhile noting that generating samples from the first module parameters is, by far, the most computationally expensive component of all the algorithms.  Since the joint sampler for the full posterior requires almost five times more iterations than the sampler for the working prior to mix properly, this means that the overall execution time for the full posterior's MCMC algorithm is substantially higher than the execution time for the cut and two-step samplers.  The execution time for the cut posterior's MCMC algorithm is clearly longer than that of the two-step posterior, but since the second-module sampler is relatively fast, the difference is relatively small.

\subsubsection{Comparing $p_f(\bfeta, \bfxi \mid \bfY)$, $p_t(\bfeta, \bfxi \mid \bfY)$ and $p_c(\bfeta, \bfxi \mid \bfY)$}\label{se:comp_xi_eta}

\begin{figure*}
\centering
\includegraphics[width=0.85\textwidth]{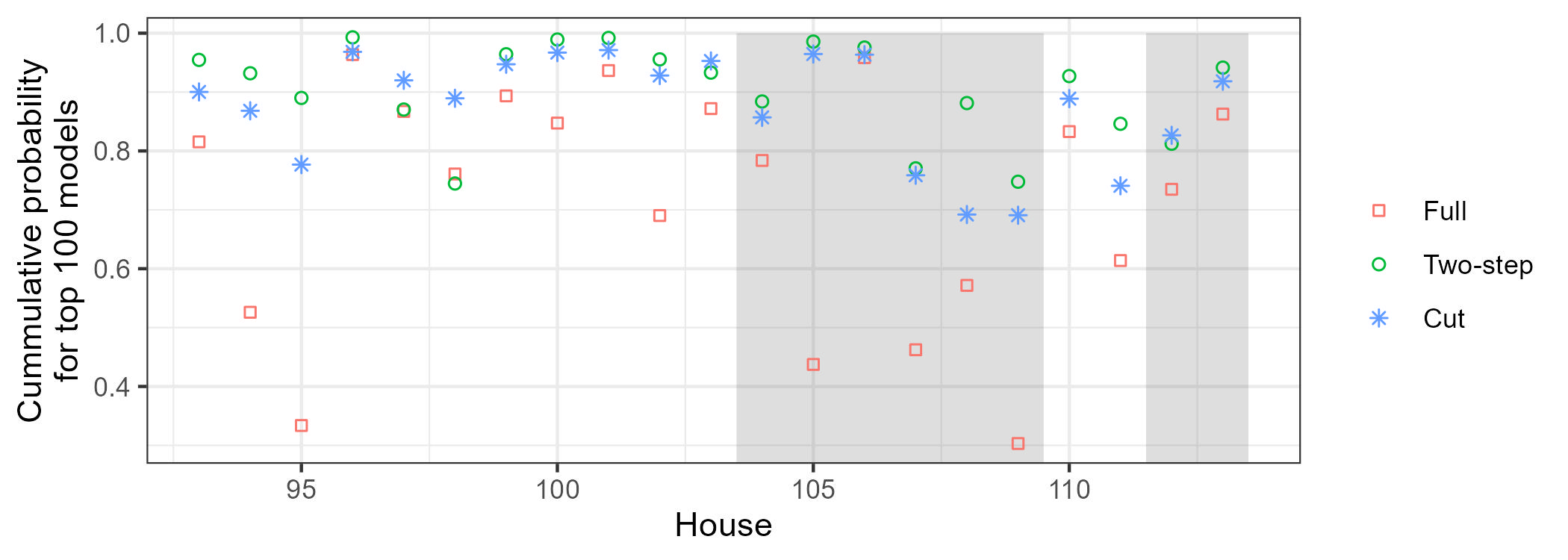}
\caption{Cumulative posterior probability for the top $100$ models $\bfxi$ with highest posterior probability. Shading indicates congresses with a Republican majority.}
\label{fig:cummprob_xi}
\end{figure*}

In this section we focus on comparing the full, cut and two-step with $a_1=a_2$ posteriors for the key parameter of interest $\bfxi$, and the associated set of coefficients $\bfeta$.  We start by presenting in Figure \ref{fig:cummprob_xi} the cumulative  probability for the $100$ models $\bfxi$ with the highest individual posterior probability for each dataset and each of the three posteriors.  We use this metric as a proxy for how concentrated each of the posterior distributions is over the model space.  We can see that, in all cases, the cumulative posterior probability for the full posterior is less than or equal to that associated with  both the two-step and the cut posteriors.  This suggests that the full posterior distribution for $\bfxi$ is less concentrated than the corresponding two-step or cut posteriors.  This is consistent with what we observed in Section \ref{se:matrixvariatelinear}.  We also note that the cut posterior often (but not always) appears to be less concentrated than the two-step posterior.  The difference with respect to what we observed in Section \ref{se:matrixvariatelinear} might be due to the non-linearities involved in this more complex illustration. 

We compare now point estimates for $\bfxi$ from each of the three posteriors.  To do so, we focus on the posterior inclusion probabilities (PIPs). The PIP for variable $k$ is defined as
\begin{equation*}
\Pr(\xi_k=1 \mid \mbox{data})=\E\left\{ \xi_k \mid \mbox{data} \right\} \approx \frac{1}{B} \sum_{b=1}^{B}  \xi_k^{(b)} .
\end{equation*}

\begin{figure*}
\centering
\includegraphics[width=0.85\textwidth]{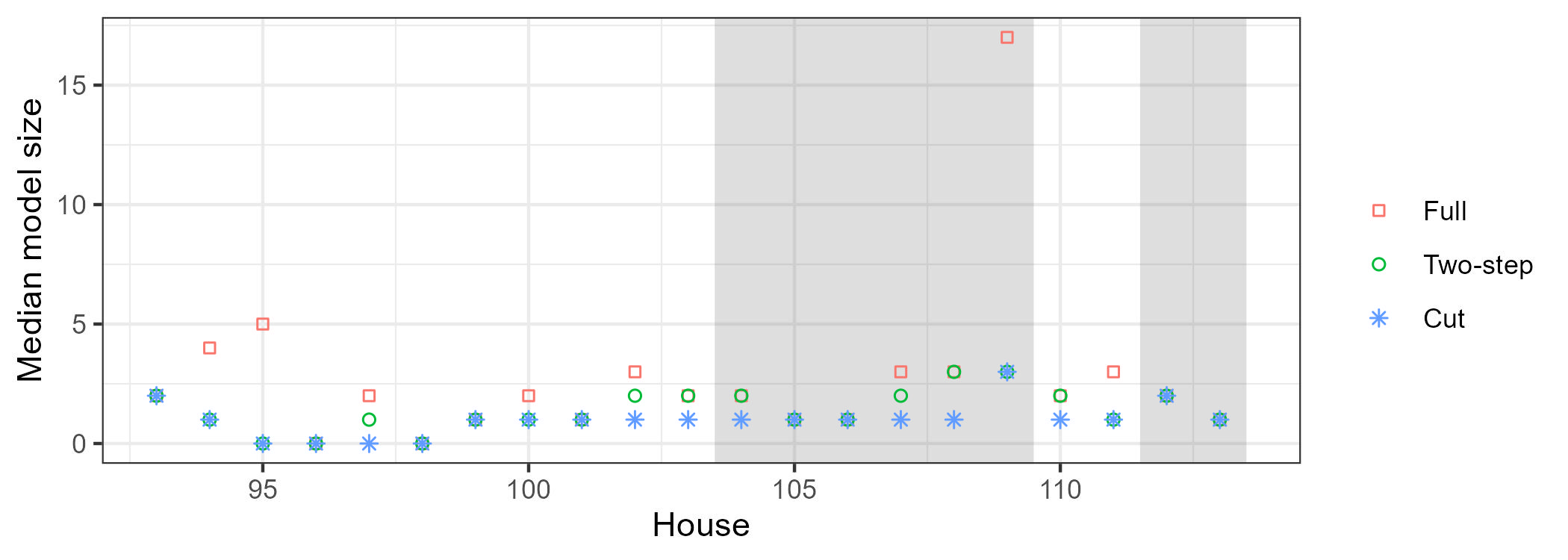}
\caption{Number of covariates included in the posterior median model for each House under consideration. Shading indicates congresses with a Republican majority.}
\label{fig:modelsize}
\end{figure*}

PIPs can be used to identify the ``median'' model, which includes those variables with a PIP greater than or equal to 0.5. Figure~\ref{fig:modelsize} shows the number of variables included in the median model by each of the three posteriors and for each of our 21 datasets.  In almost all cases, the median model includes between $0$ and $5$ covariates. The  exception is the full posterior for the 109\ts{th} House, which includes 17 out of the 21 variables in the median model.  This a surprisingly large number, especially when considering that the cut and two-step posteriors each include only 3 variables, and suggests a possible model misspecification issue. 
It is also worthwhile noting that, across all datasets, the full posterior includes at least as many variables as the two-step posterior, which in turn includes at least as many as the cut posterior.

\begin{figure*}
  \centering
  \includegraphics[width=\linewidth]{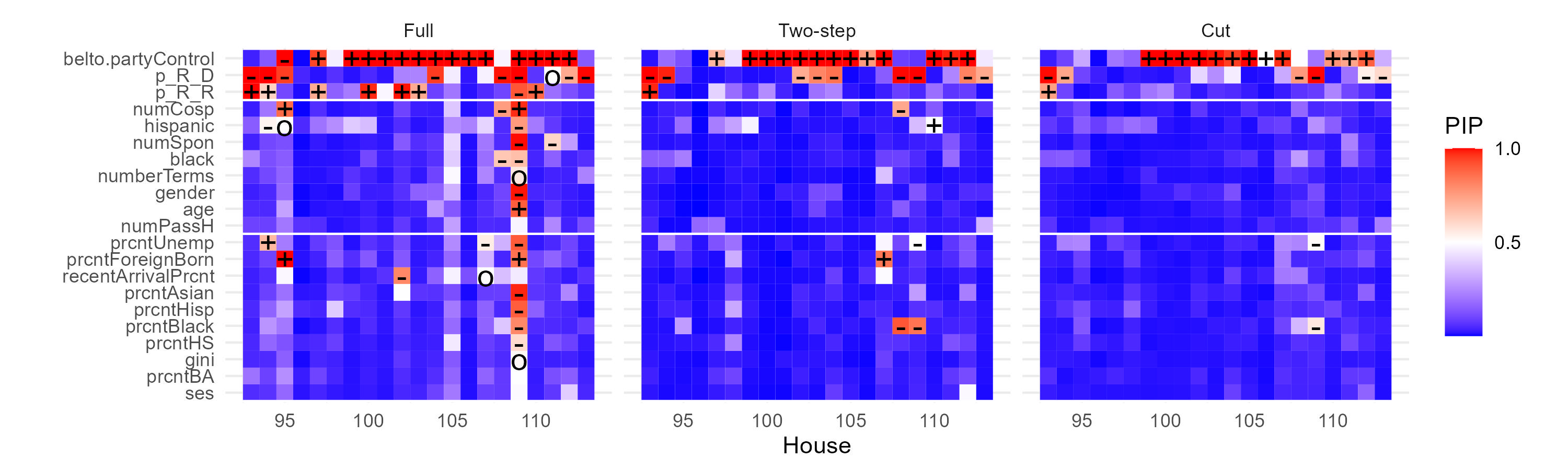}
  \label{fig:PIPhearmap_fullpost}
    \caption{Heatmaps showing posterior inclusion probabilities under each of the three posteriors under consideration.}
    \label{fig:heatmaps}
\end{figure*}

\begin{figure*}
\centering
\includegraphics[width=0.85\textwidth]{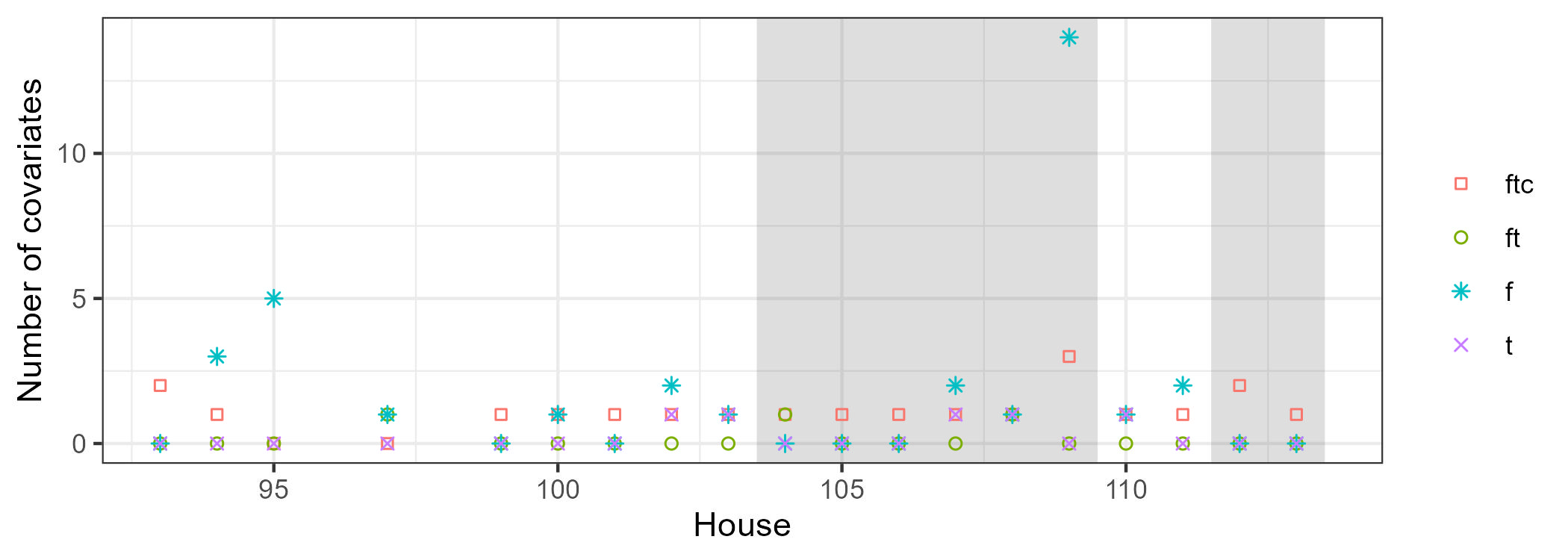}
\caption{Breakdown of the number of covariates included in different sets of models (full (f), two-step (t), and cut (c)). Shading indicates congresses with a Republican majority.}
\label{fig:modelsizebreakdown}
\end{figure*}

To gain further insight into these posteriors for $\bfxi$, we present in Figure \ref{fig:heatmaps} heatmaps for the posterior inclusion probabilities and in Figure \ref{fig:modelsizebreakdown} a breakdown of how many variables are selected by each model on its own and by any combination of them.  It is notable that the variables included in the median model by the cut posterior are a subset of the variables included by the two-step and full posteriors.  In most cases, the variables selected by the two-step posteriors are also a subset of those included by the full posterior.  However, there are a few exceptions.  For example in the 102\textsuperscript{nd} and 103\textsuperscript{rd} Houses, the median model under the two-step posterior includes the proportion of constituency-level Republican vote in the most recent election for Democrat legislators, but not the proportion of constituency-level Republican vote in the most recent election for Republican legislators.  This is the mirror image of what the full model does for these Houses.  It is also worthwhile noting that, for the 109\textsuperscript{th} House, the two-step and cut posteriors select the same three variables.  This is again consistent with what we saw in Section \ref{se:matrixvariatelinear} in terms of robustness to model misspecification.

\begin{figure*}
  \begin{subfigure}[b]{.9\textwidth}
  \centering
  \includegraphics[width=\textwidth]{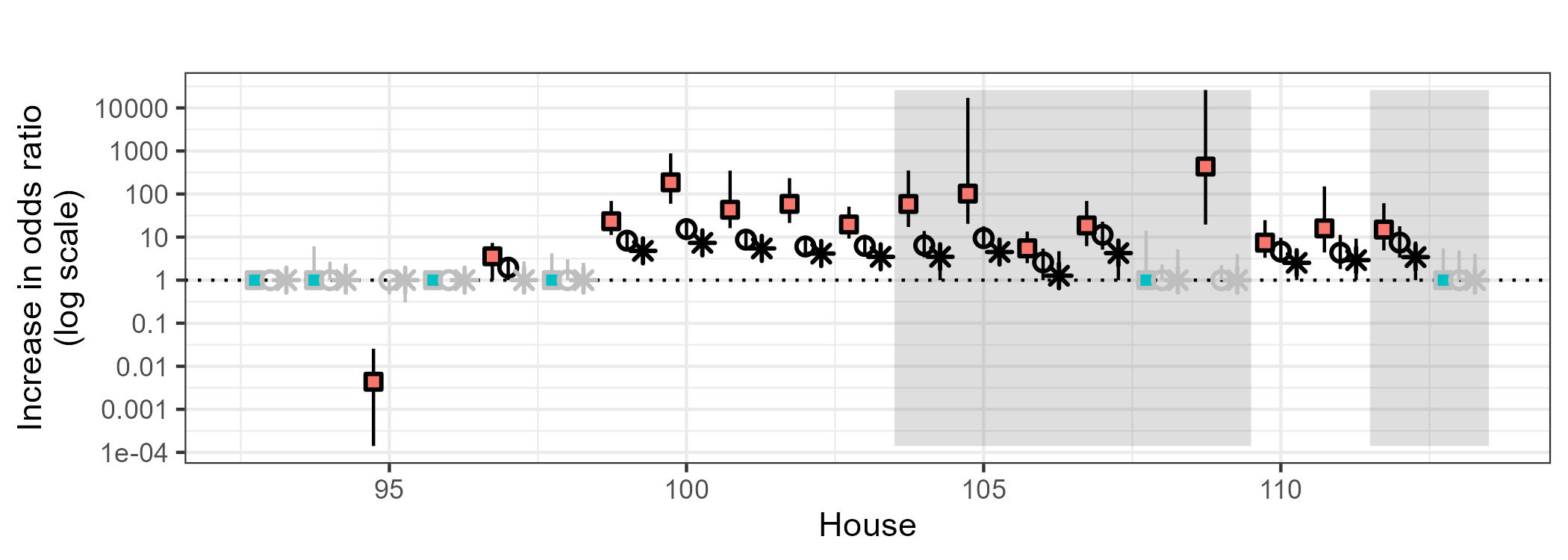}
  \caption{Influence of belonging to the majority party on odds of being a bridge legislator}\label{fig:beltoparty}
  \end{subfigure}
  \begin{subfigure}[b]{.9\textwidth}
  \centering
  \includegraphics[width=\textwidth]{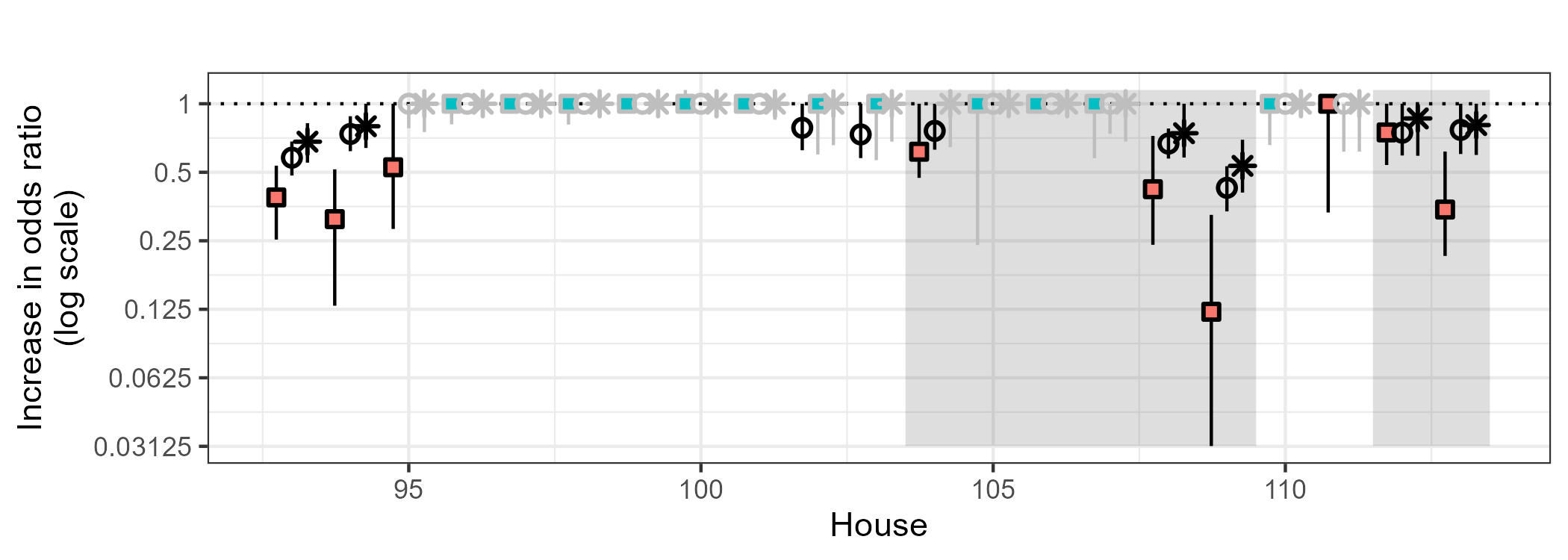}
  \caption{Influence of proportion of constituency-level Republican vote in the most recent presidential election on odds of being a bridge legislator, for Democrats only}\label{fig:prD}
  \end{subfigure}
  \begin{subfigure}[b]{.9\textwidth}
  \centering
  \includegraphics[width=\textwidth]{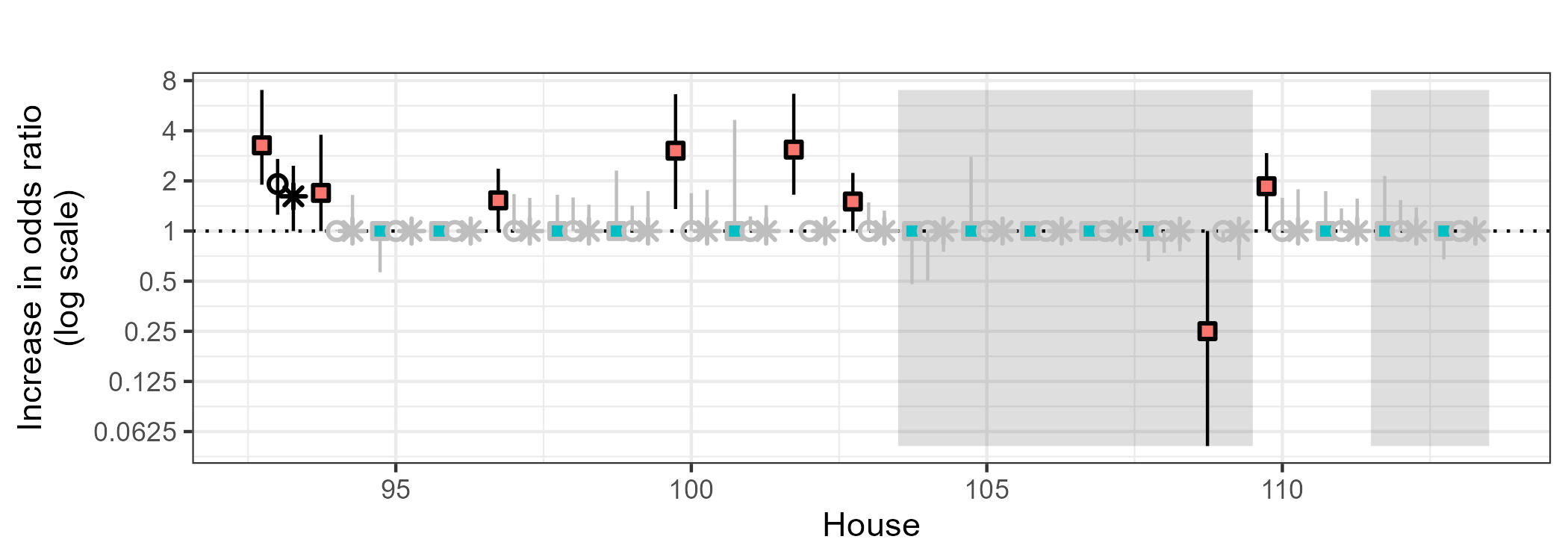}
  \caption{Influence of proportion of constituency-level Republican vote in the most recent presidential election on odds of being a bridge legislator, for Republicans only.}\label{fig:prR}
  \end{subfigure}
\caption{Posterior mean and 95\% percent credible intervals for the increase in odds ratios of being a bridge legislator for the three most important variables identified by our analysis. Squares ($\square$) represent the full model, circles ($\circ$) represent the two-step posterior, and asterisks ($\ast$) represent the cut posterior. Black lines indicate that the covariate was selected into the model ($PIP\geq0.5$) and gray lines indicate that the covariate was not selected.}
\label{fig:covariates}
\end{figure*}

To conclude this section, we focus on the posterior distribution of the regression coefficients $\bfeta$ under each of the three posterior distributions. We focus the discussion on what appears to be the three most important variables:  whether the legislator belongs to the majority party in the House, and the interactions between the proportion of constituency-level Republican vote and legislator's party membership indicator.  The most commonly selected variable is the indicator for whether a legislator belongs to the party that holds the majority of seats for that session of the House. This variable is significant in 15, 13, and 12 of the Houses under study respectively for the full, two-step, and cut posteriors. Figure~\ref{fig:beltoparty} displays the increase in the odds of being a bridge associated with being in the majority party. The three posteriors are consistent in, for the most part, estimating a positive association between being in the majority party and being a bridge.  This result indicates that legislators in the majority are more like to exhibit the same preference when voting on procedural and final passage votes. In a similar trend to model size, the estimates tend to be largest (in absolute value) for the full posterior, followed by the two-step and then the cut posteriors.  In some cases, these larger estimates appear to be excessively large under the full posterior (for example the ratio is greater than $100$ for the $100$\ts{th}, $105$\ts{th}, and $109$\ts{th} Houses), again suggesting possible misspecification issues.

The next two most commonly selected variables are related to the political leanings of the district represented by the legislator, as measured by the share of the district's two-party vote gathered by the Republican presidential candidate in the most recent presidential election. We have observed that the distribution of this covariate has different dynamics depending on the party affiliation of the legislator representing the district, and so we examine the interaction of this variable with party affiliation. Figures~\ref{fig:prD} and \ref{fig:prR} display the increase in the odds of being a bridge corresponding to a $5$ percent increase in the Republican vote share, for Democratic and Republican legislators respectively. For the interaction related to Democratic legislators, the variable is significant in $9$ Houses for the full and two-step approaches and $6$ Houses for the cut approach. The sign of this association is generally negative, meaning that Democratic legislators with more moderate constituencies are less likely to be bridges. For the two-step and cut approaches, a $5\%$ decrease in the Republican vote share for the district is is associated with an increase of $1.12$ to $2.34$ in the bridging probability. In contrast the estimates for the full model are as high as $8.2$, with the most extreme estimate again relating to the outlier $109$\ts{th} House. The interaction associated with the Republican party is significant under the full model in $6$ Houses, but it is significant for only $1$ House under the two-step and cut approaches. The sign of this association is generally positive meaning that Republican legislators with more moderate constituencies are also less likely to be bridges. As before, the estimate for the $109$\ts{th} House is an outlier, this time in both sign and magnitude.

\subsubsection{Sensitivity of the two-step posterior to the underlying utility function}

One of the shortcomings of two-step procedures is their dependence on the utility function used to construct the point estimators from the first module.  In this particular case study, the issue is what value to use to threshold $\bar{p}(\zeta_i = 1 \mid \bfY)$.

\begin{figure*}
\centering
\includegraphics[width=0.85\textwidth]{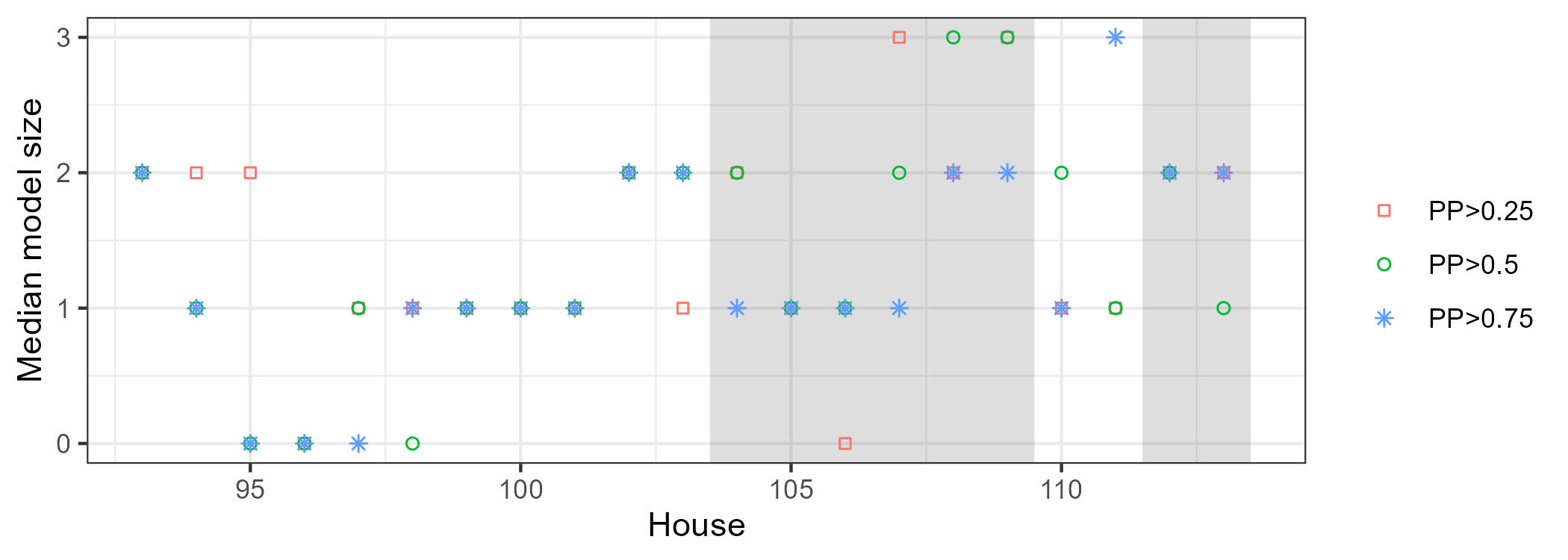}
\caption{Number of covariates included in the posterior median model for each House under consideration under the two-step posterior for different thresholds. Shading indicates congresses with a Republican majority.}
\label{fig:modelsize_sens}
\end{figure*}

Figure \ref{fig:modelsize_sens} shows the number of covariates included in the median model for the two-step posterior constructed using three different values of $a_1/(a_1 + a_2)$ (recall Section \ref{se:mcmc_twostep}):  0.5 (our default, which was used in all comparisons presented in Section \ref{se:comp_xi_eta}), 0.25 (which favors a larger number bridges) and 0.75 (which favors fewer bridges). Because the number of significant variables tends to be small in general, the impact of the threshold in this case study is moderate.  Nonetheless, we can see that the threshold does affect which variables are selected, and that there is no obvious pattern:  a higher threshold might lead either to fewer or additional variables in the model depending on the dataset at hand.

\begin{figure*}
  \begin{subfigure}[b]{.9\textwidth}
  \centering
  \includegraphics[width=\textwidth]{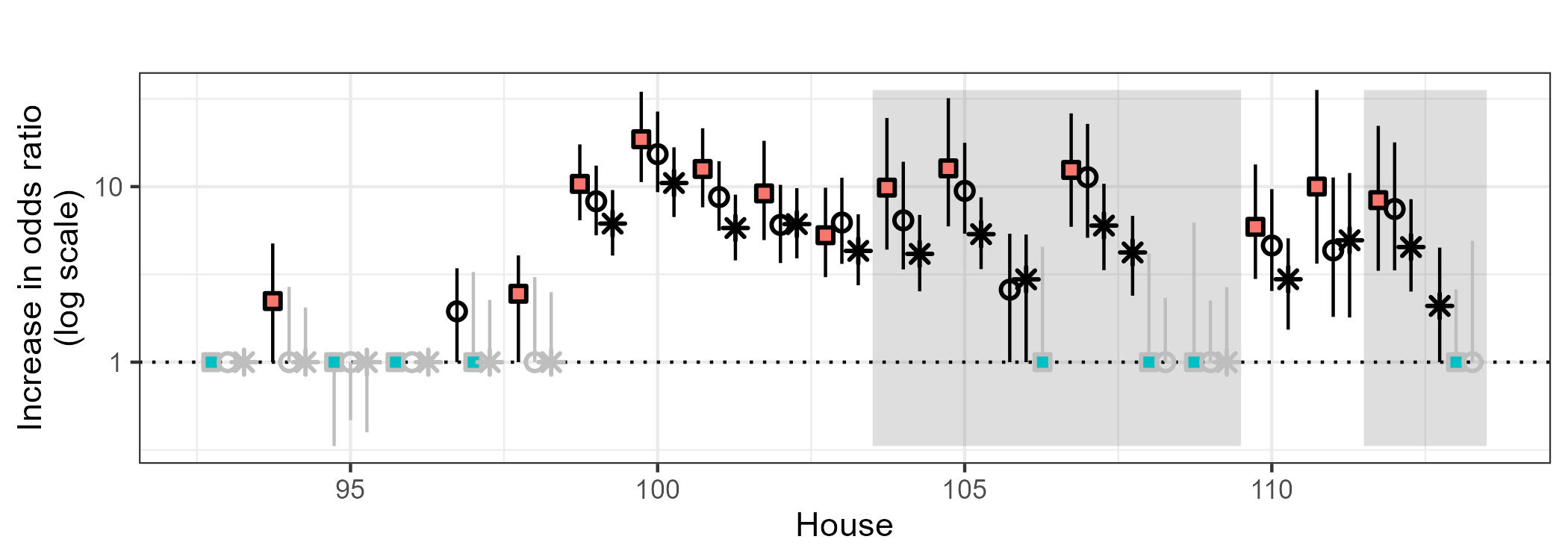}
  \caption{Influence of belonging to the majority party on odds of being a bridge legislator}\label{fig:beltoparty_sens}
  \end{subfigure}
  \begin{subfigure}[b]{.9\textwidth}
  \centering
  \includegraphics[width=\textwidth]{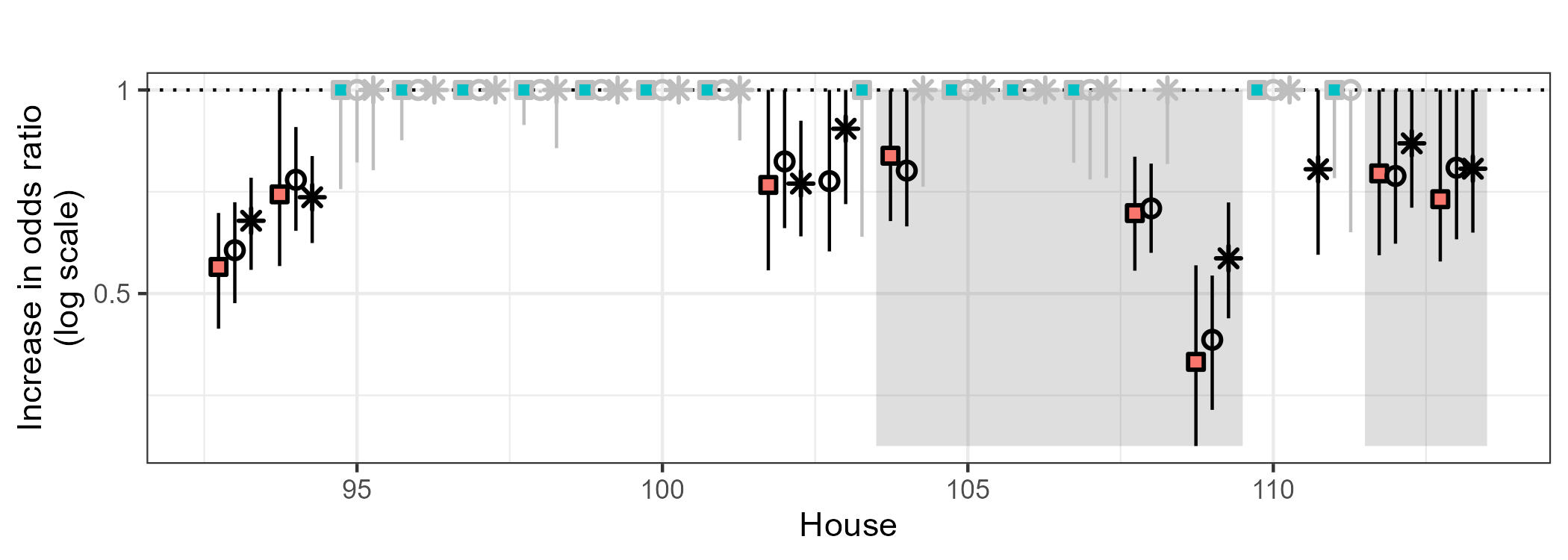}
  \caption{Influence of proportion of constituency-level Republican vote in the most recent presidential election on odds of being a bridge legislator, for Democrats only}\label{fig:prD_sens}
  \end{subfigure}
  \begin{subfigure}[b]{.9\textwidth}
  \centering
  \includegraphics[width=\textwidth]{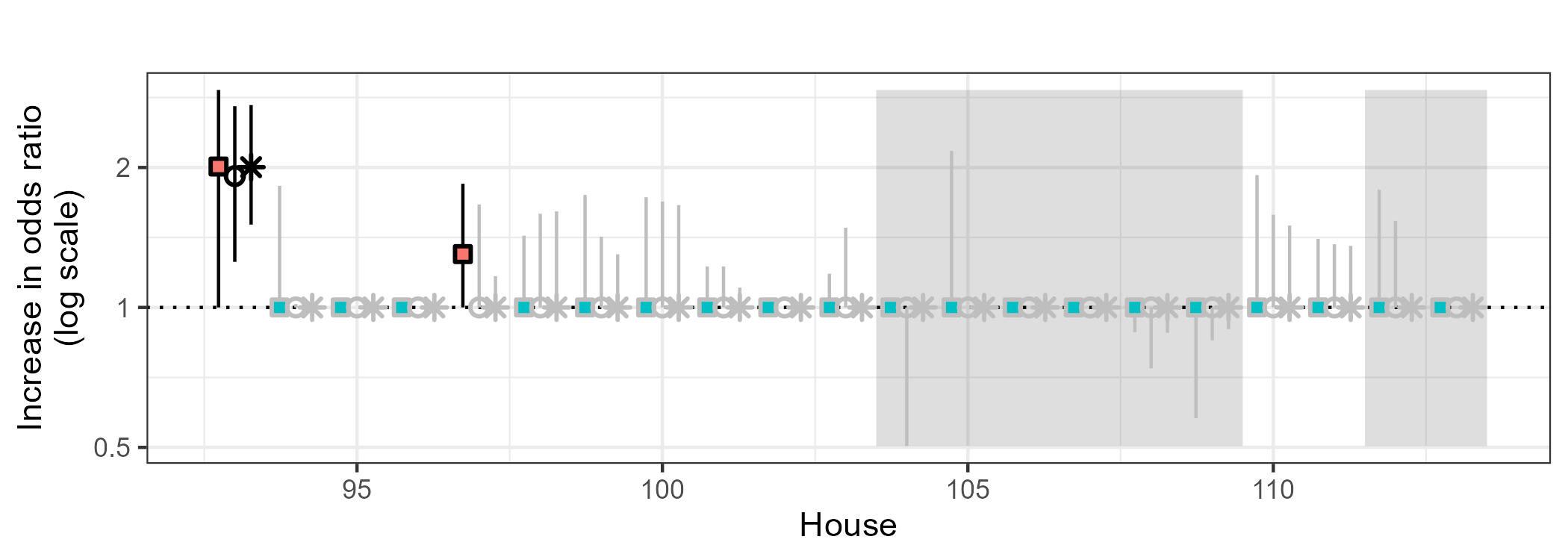}
  \caption{Influence of proportion of constituency-level Republican vote in the most recent presidential election on odds of being a bridge legislator, for Republicans only.}\label{fig:prR_sens}
  \end{subfigure}
\caption{Posterior mean and 95\% percent credible intervals for the increase in odds ratios of being a bridge legislator, under the two-step posterior for different thresholds, for the three most important variables identified by our analysis. Squares ($\square$) represent a threshold of $PP>0.25$, circles ($\circ$) represent a threshold of $PP>0.5$, and asterisks ($\ast$) represent a threshold of $PP>0.75$. Black lines indicate that the covariate was selected into the model ($PIP\geq0.5$) and gray lines indicate that the covariate was not selected.}
\label{fig:covariates_sens}
\end{figure*}

We consider now the effect of the threshold on the estimates of $\bfeta$.  As in Section \ref{se:comp_xi_eta}, we focus on three variables:  whether the legislator belongs to the majority party in the House, and the interactions between the proportion of constituency-level Republican vote and legislator's party membership indicator.  Figure \ref{fig:covariates_sens} presents the posterior mean and corresponding 95\% credible intervals for the increase in the odds of being a bridge associated with each of the three variables for the two-step posterior associated with each of the three thresholds.  Overall, when the three posteriors agree in including a given variable in the model, the effect of the threshold on the estimates of the coefficients is relatively moderate.  Note, however, the difference in vertical scales between Figure \ref{fig:beltoparty} and Figure \ref{fig:beltoparty_sens}: while all versions of the two-step posterior tend to generate effects that are of the same order of magnitude, these can be a lot smaller than those generated by the full model.

\subsubsection{Comparing $p(\bfzeta \mid \bfY)$ and $\bar{p}(\bfzeta \mid \bfY)$}

\begin{figure*}
\centering
\includegraphics[width=0.85\textwidth]{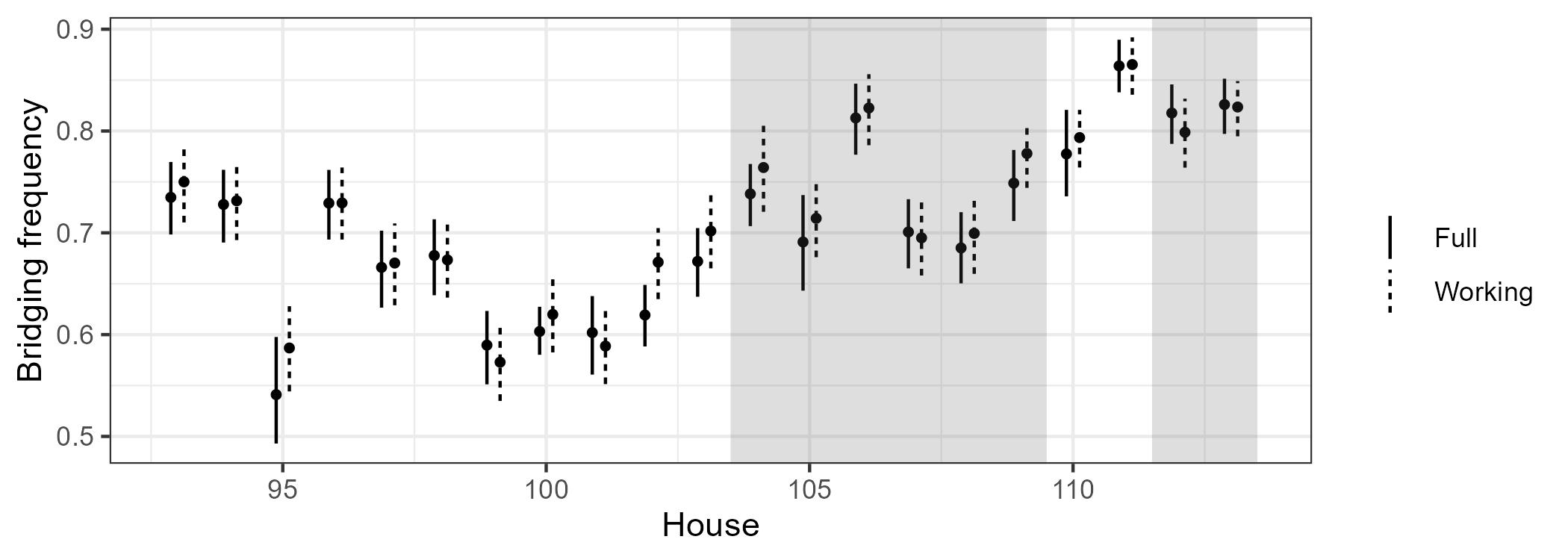}
\caption{Posterior mean and 95\% credible intervals for the bridging frequency estimated by the full and working first-level posteriors, $p(\bfzeta \mid \bfY)$ and $p(\bar{\bfzeta} \mid \bfY)$.}
\label{fig:abf}
\end{figure*}

While inferences on $\bfzeta$ are not the focus of this application, we nonetheless explore the difference between the full and working posteriors in this context.  To accomplish this we focus on the Bridging Frequency (BF), $BF = \frac{1}{N}\sum_{i=1}^N \zeta_i$, which is an aggregate metric of coherence across voting domains that can be useful for descriptive purposes.  Figure \ref{fig:abf} presents the posterior mean and 95\% credible intervals for the bridging frequency estimated by the full and working first-level posteriors.  The biggest differences seem to arise in the 95\textsuperscript{th} and 102\textsuperscript{th} to 104\textsuperscript{th} Houses.  However, the overall behavior over time is very similar and there is substantial overlap between the posterior credible intervals. 


\section{Discussion}\label{sec:discussion}

While data analysis pipelines are widely used in practice, there are substantial challenges with their use that are often underappreciated.  From a Bayesian perspective, it is tempting to argue that integrating the various modules of the pipeline into a single hierarchical model to be jointly fitted to the data is the solution to most of them.  However, implementing such big hierarchical models might involve substantial computational challenges, both in terms of algorithm development and of execution time.  More importantly, as our two illustrations show, hierarchical Bayesian models are particularly susceptible to model misspecification.  The results from  our second illustration provide a particularly sharp warning:  the full posterior seems to perform quite well for most datasets, but appears to fail dramatically for the 109\textsuperscript{th} House, and perhaps for the 100\textsuperscript{th} and 105\textsuperscript{th} Houses as well.  The best solution to this problem, investigating the sources of misspecification and modifying the model accordingly, can be laborious  and, in some cases, unfeasible.  Our illustrations show that cut inference provides an alternative to full hierarchical inference that can be more robust to model misspecification, while being better at reflecting the true level of uncertainty in the estimates than more traditional multi-step approaches to statistical inference.

Most of the literature on cut inference has focused on models where there are two separate sets of response variables that are assumed to be conditionally independent given a set of shared parameters.  As far as we are aware, this paper is the first one to explore cut inference in sequential settings such as those associated with data analysis pipelines.  However, much work remains to be done.  The result in Theorem \ref{lem:KL} is useful in terms of establishing the optimality of cut inference among procedures that share a given marginal distribution for $\bfzeta$ (including multi-step procedures), and the ideas underlying the derivation of Equation \eqref{eq:klbound} can be extended to produce bounds for the Kullback–Leibler divergence between the cut and full posteriors in settings where the working prior $\bar{p}(\bfzeta)$ is carefully chosen to be compatible with the original $p(\bfzeta \mid \bfxi)$.  However, it would be useful to have tools that would allow us to accurately quantify the impact of cut inference on parameter uncertainty.  Exploring potential connections between cut inference and (constrained) variational inference (e.g., see \citealp{blei2017variational}) might provide a path forward in this. 


\begin{appendix}

\section{Hyperpriors for Illustration 2}\label{ap:model_description1}

The definition of the first module (Section~\ref{se:module1}) is completed by specifying priors on the parameters $\bfmu=(\mu_1,\dots,\mu_J)$ and $\bfalp=(\alf_1,\dots,\alf_J)$, as well as the hyperparameters $\bfrho_{\beta}$ and  $\sig^2_{\beta}$.  The hyperparameter $\bfrho_{\beta}$ in Equation~\eqref{eq:priorbeta} is given a standard Gaussian hyperprior. The hyperparameter $\sig^2_{\beta}$ from the same equation is given an inverse gamma hyperpriors with shape parameter $2$ and scale parameter $1$ (for a mean of $1$ and an infinite variance).  On the other hand, the prior on the intercept $\bfmu$ from Equation~\eqref{eq:likelihood2} is given by
\begin{align*}
\mu_j \mid \rho_\mu, \kap^2_\mu & \simiid \normal(\mu_j \mid \rho_\mu, \kap^2_\mu), & j&=1, \ldots, J,
\end{align*}
where the hyperparameters $\rho_\mu$ and $\kap^2_\mu$ are then given as hyperpriors a standard normal and inverse gamma with shape parameter 2 and scale parameter 1,  respectively. For the discrimination parameter $\bfalp$ in Equation~\eqref{eq:likelihood2} we set 
\begin{align*}
\alf_{j} \mid \omega_\alf, \kap^2_\alf  &\simiid \omega_{\alf} \del_{0}(\alf_{j}) 
+ (1-\omega_{\alf})\normal(\alf_{j} \mid 0,\kap^2_\alf),
\end{align*}
for $j=1,\ldots, J$,  where $\delta_0$ is a point mass at $0$. This prior is completed by assigning an inverse gamma hyperprior with shape parameter 2 and scale parameter 1 on $\kappa_{\alpha}^2$, and a uniform prior on $\omega_{\alf}$.


\section{Explanatory variables used on Illustration 2}\label{ap:model_description2}

Demographic and socioeconomic data for the legislators and constituencies was obtained from \cite{foster2017historical}.  Data on the results of the presidential elections between 1970 and 2008 was generously provided by Stephen Jessee and Sean Theriault (personal communication), while data for the 2012 and 2016 elections was obtained from 
\cite{KosReport}.

\subsection{Covariates related to party affiliation}

\begin{description}
\item[belto.partyControl:] Indicator for whether the legislator is a member of the party that has the majority in the current House.
\item[p\_R\_R and p\_R\_D:] These two covariates are interactions between the two party membership indicators and a measure of partisan political ideology for the legislator's district. The Republican share of the two-party presidential vote (p\_R) is the percentage of the two-party vote won by the Republican candidate in the most recent presidential election (centered so that $0$ indicates that the two parties received an equal percentage of the vote). $p\_R\_R=I(Republican)\times p\_R$ is equal to the Republican voteshare $p\_R$ for legislators belonging to the Republican party and $0$ for legislators belonging to the Democratic party. The other interaction $p\_R\_D=I(Democrat)\times p\_R$ is defined an analogous way. 
\end{description}

\subsection{Legislator characteristics}
\begin{description}
\item[age:] Age at time of being sworn into congress for the current session.
\item[gender:] Gender of legislator.
\item[black:] Indicator for membership to the Congressional Black Caucus. The authors of the data note that to their knowledge, all self-identifying Black members of congress are members of the caucus.
\item[hispanic:] Indicator for membership to the Congressional Hispanic Caucus. The authors of the data note that to their knowledge, all self-identifying Hispanic members of congress are members of the caucus.
\item[numberTerms:] Number of terms served in the House.
\item[numSpon:] Number of bills sponsored by the legislator in the current term. 
\item[numCosp:] Number of bills co-sponsored by the legislator in the current term.
\item[numPassH:] Number of bills sponsored by the legislator that were approved by a full House vote in the current term.
\end{description}

\subsection{Constituency characteristics (based on data from Census)}

\begin{description}
\item[recentArrivalPrcnt:] Percentage of the district that recently moved to the district from another county (note that the census does not track how many people have moved into a district from within the same county).
\item[prcntForeignBorn:] Percentage of the district that was born in a foreign country.
\item[gini:] Index of economic inequality calculated based on the percentage of the population in each income bracket.
\item[ses:] Measure of socioeconomic status calculated based on the income and education level of the district. 
\item[prcntUnemp:] Percentage of the district's population that is unemployed but still in the labor force.
\item[prcntBA:] Percentage of the district with a bachelor's degree or higher.
\item[prcntHS:] Percentage of the district with a high school degree or higher.
\item[prcntBlack:] Percentage of the district that is Black, including those who are Black and Hispanic.
\item[prcntHisp:] Percentage of the district that is Hispanic (both Black and White).
\item[prcntAsian:] Percentage of the district that is Asian.
\end{description}

\end{appendix}

\begin{acks}[Acknowledgments]
We would like to thank Stephen Jessee and Sean Theriault for providing access to key data for our empirical illustrations.
\end{acks}
\begin{funding}
This research was partially supported by NSF grants NSF-2023495 and NSF-2114727.
\end{funding}

\bibliographystyle{imsart-nameyear}
\bibliography{advancementbib, newrefs}       


\end{document}


\title{Supplementary materials for ``On Data Analysis Pipelines and Modular Bayesian Modeling''}
\author{Erin Lipman and Abel Rodriguez \\ University of Washington, Seattle}
\maketitle

\section{Proof of Lemma 1}\label{ap:mainlemmaproof}
\begin{proof}
    First note that for any $q\in\F$,
    \begin{align*}
        \KL{q(\bfzeta,\bfxi)}{p_f(\bfzeta,\bfxi\mid \bfY)} &=
        \E_q\left[log
        \frac{q(\bfzeta,\bfxi)}{p_f(\bfzeta,\bfxi\mid \bfY)}\right] \\
        &=\E_q\left[log
        \frac{q(\bfxi\mid\bfzeta)}{p_f(\bfxi\mid \bfzeta)}\right] +
        \E_q\left[log
        \frac{\bar{p}(\bfzeta\mid \bfY))}{p_f(\bfzeta\mid \bfY)}\right] \\
        &=\E_{q_{\bfzeta}}\E_{q_{\bfxi\mid\bfzeta}}\left[log
        \frac{q(\bfxi\mid\bfzeta)}{p_f(\bfxi\mid \bfzeta)}\right] +
        \E_{q_{\bfzeta}}\E_{q_{\bfxi\mid\bfzeta}}\left[log
        \frac{\bar{p}(\bfzeta\mid \bfY)}{p_f(\bfzeta\mid \bfY)}\right] \\
        &=\E_{q_{\bfxi\mid\bfzeta}}\left[log
        \frac{q(\bfxi\mid\bfzeta)}{p_f(\bfxi\mid \bfzeta)}\right] +
        \E_{q_{\bfzeta}}\left[log
        \frac{\bar{p}(\bfzeta\mid \bfY)}{p_f(\bfzeta\mid \bfY)}\right] \\
        &=\KL{q(\bfxi\mid\bfzeta)}{p_f(\bfxi\mid \bfzeta)}
        +\KL{\bar{p}(\bfzeta\mid \bfY)}{p_f(\bfzeta\mid \bfY)}.
    \end{align*}
    The second term in the sum is constant in $q$ on the class $\F$, and thus the expression is minimized when the first term is $0$, that is $q(\bfxi\mid\bfzeta)=P_f(\bfxi\mid\bfzeta)$. This gives
    \[\argmin\limits_{q\in \F}
        \KL{q(\bfzeta,\bfxi)}{p_f(\bfzeta,\bfxi\mid \bfY)}
    =p_f(\bfxi\mid\bfzeta)\cdot \bar{p}(\bfzeta\mid \bfY)=p_c(\bfzeta,\bfxi\mid\bfY).\]
    Plugging in $q(\bfzeta,\bfxi)=p_c(\bfzeta,\bfxi\mid\bfY)$ to the final expression in the above derivation gives the second part of the result.
\end{proof}

To apply Lemma 1 to derive the bound in Equation 18, we observe that
\[
\frac{\bar p(\bfzeta\mid\bfY)}{p_f(\bfzeta\mid\bfY)}
=\frac{p_f(\bfzeta\mid\bfY,\bfxi=\zk)}{p_f(\bfzeta\mid\bfY)}
=\frac{p_f(\bfxi=\zk\mid\bfzeta)}{p_f(\bfxi=\zk\mid\bfY)}
\leq \frac{1}{p_f(\bfxi=\zk\mid \bfY)}.
\]

\section{Derivation of the posterior distributions for the two-step matrix linear posterior}\label{ap:matrixlinemodel}

Recall that the density associated with a matrix variate normal distribution $\mvn_{N,J} ( \mathbf{M}, \mathbf{U}, \mathbf{V})$ is
\begin{align*}
\frac{\exp\left\{ -\frac{1}{2} \tr \left[ \mathbf{V}^{-1} \left( \bfY - \mathbf{M} \right)^T \mathbf{U}^{-1} \left( \bfY - \mathbf{M} \right) \right]\right\}}{(2\pi)^{NJ/2} \left| \mathbf{U} \right|^{J/2} \left| \mathbf{V}\right|^{N/2}}  .
\end{align*}

Two interrelated results about convolutions of matrix variate Gaussian distributions will be important in our derivations.  First, if $\bfY_1 \sim \mvn_{N,J}(\bfM_1, \bfU_1, \bfV)$ and $\bfY_2 \sim \mvn_{N,J}(\bfM_2, \bfU_2, \bfV)$, then $\bfY_1 + \bfY_2 \sim \mvn_{N,J}(\bfM_1 + \bfM_2, \bfU_2 + \bfU_2, \bfV)$.  Similarly, if $\bfY_1 \sim \mvn_{N,J}(\bfM_1, \bfU, \bfV_1)$ and $\bfY_2 \sim \mvn_{N,J}(\bfM_2, \bfU, \bfV_2)$, then $\bfY_1 + \bfY_2 \sim \mvn_{N,J}(\bfM_1 + \bfM_2, \bfU, \bfV_1 + \bfV_2)$.  Another important pair of results involves linear transformations of matrix variate normal distributions.  If $\mathbf{D}$ and $\mathbf{C}$ are matrix $N \times r$ and $J \times s$ matrices, and $\bfY \sim \mvn_{N,J}(\bfM, \bfU, \bfV)$, then $\mathbf{D} 
\bfY \mathbf{C} \sim \mvn_{N,J}(\mathbf{D}\bfM\mathbf{C} , \mathbf{D}\bfU\mathbf{D}^T, \mathbf{C}^T \bfV \mathbf{C})$.  In our first illustration, these results can be used to derive $p(\bfY \mid \bfxi) = \int p(\bfY \mid \bfzeta) p(\bfzeta \mid \bfxi) \dd \bfzeta$ by noting that $\bfY = (\bfX \bfxi + \bfEta) \bfW + \bfEpsilon = \bfX \bfxi \bfW + \bfEta\bfW + \bfEpsilon$, where $\bfEta \bfW \sim \mvn_{N,J} (\mathbf{0}, \mathbf{I}_N, \tau^2 \bfW^T\bfW)$ and therefore $\bfEta \bfW  + \bfEpsilon \sim \mvn_{N,J}(\mathbf{0}, \mathbf{I}_N,\tau^2 \bfW^T\bfW + \sigma^2 \mathbf{I}_{J})$.

\subsection{Full posterior}\label{ap:derivationfullpost}

Letting $\bfOmega = \tau^2 \bfW^T\bfW + \sigma^2 \mathbf{I}_{J}$, we have
\begin{align*}
    p_f(\bfxi \mid \bfY) &\propto \exp \left\{ -\frac{1}{2} \tr \left[\bfOmega^{-1} \left( \bfY - \bfX\bfxi\bfW \right)^T \left( \bfY - \bfX\bfxi\bfW\right)  \right]\right\} \\
    &\propto \exp \left\{ -\frac{1}{2} \left( \tr \left[ \bfOmega^{-1}  \bfW^T \bfxi^T \bfX^T \bfX\bfxi\bfW \right] - \tr \left[ \bfOmega^{-1}\bfY^T  \bfX\bfxi\bfW\right] - \tr \left[ \bfOmega^{-1} \bfW^T \bfxi^T \bfX^T \bfY \right]\right)
    \right\} \\
    &= \exp \left\{ -\frac{1}{2} \left( \tr \left[ \bfW\bfOmega^{-1}  \bfW^T \bfxi^T \bfX^T \bfX\bfxi \right] - 2\tr \left[ \bfW\bfOmega^{-1}\bfY^T  \bfX\bfxi\right] \right)
    \right\} \\
    &\propto \exp \left\{ -\frac{1}{2}  \tr \left[ \bfW\bfOmega^{-1}  \bfW^T \left(\bfxi - \hat{\bfxi}(\bfY) \right)^T \bfX^T\bfX \left(\bfxi - \hat{\bfxi}(\bfY)\right) \right]
    \right\},
\end{align*}
where $\hat{\bfxi}(\bfY) = [\bfX^T\bfX]^{-1}\bfX^T\bfY \bfOmega^{-1}\bfW^T [\bfW\bfOmega^{-1}  \bfW^T]^{-1}$.  The second to last equality relies on the cyclic property of the trace, while the last equality comes from a simple completion of squares in matrix form. 
Finally, note that we can apply the Woodbury inversion formula twice to get $[\bfW\bfOmega^{-1}  \bfW^T]^{-1} = \sigma^2 [\bfW\bfW^T]^{-1} + \tau^2\mathbf{I}_L$.  This implies that
\begin{align*}
p_f\left(\bfxi \mid \bfY  \right) = \mvn_{K,L} \left( \hat{\bfxi}(\bfY), [\bfX^{T} \bfX]^{-1}, \sigma^2 [\bfW \bfW^T]^{-1} + \tau^2 \mathbf{I}_{L} \right) . 
\end{align*}

To see that $\hat{\bfxi}(\bfY)$ is unbiased, note that 
\begin{align*}
    \E \left\{ \hat{\bfxi}(\bfY) \right\} &= [\bfX^T\bfX]^{-1}\bfX^T \E \left\{\bfY \right\} \bfOmega^{-1}  \bfW^T [\bfW\bfOmega^{-1}  \bfW^T]^{-1} \\
    &= [\bfX^T\bfX]^{-1}\bfX^T \bfX \bfxi\bfW \bfOmega^{-1} \bfW^T [\bfW\bfOmega^{-1}  \bfW^T]^{-1} = \bfxi . 
\end{align*}

More generally, 
\begin{align*}
    \hat{\bfxi}(\bfY)  &\sim \mvn \left(  \bfxi, [\bfX^T\bfX]^{-1}, [\bfW\bfOmega^{-1}  \bfW^T]^{-1}
    \bfW \bfOmega^{-1}
    \bfOmega_{*}
    \bfOmega^{-1} \bfW^T 
    [\bfW\bfOmega^{-1}  \bfW^T]^{-1}
    \right)
\end{align*}
where $\bfOmega_{*} = \tau^{2}_{*} \bfW^T\bfW + \sigma^{2}_{*} \mathbf{I}_{J}$ and $ \tau^{2}_{*}$ and  $\sigma^{2}_{*}$ are the  values of $\tau^2$ and $\sigma^2$ under the true data generation mechanism.

\subsection{Two-step posterior}\label{ap:derivationtwosteppost}

The calculations required to compute $p(\bfzeta \mid \bfY)$ and $p(\bfxi \mid \bfzeta)$ are very similar:
\begin{align*}
    p(\bfzeta \mid \bfY) &\propto \exp \left\{ -\frac{1}{2\sigma^2} \tr  \left[ \left( \bfY - \bfzeta \bfW \right)^T \left( \bfY - \bfzeta \bfW \right) \right] \right\} \\
    & \propto \exp \left\{ -\frac{1}{2\sigma^2}\left( \tr \left[\bfW^T\bfzeta^T\bfzeta\bfW \right]
    - \tr \left[\bfY^T\bfzeta\bfW \right]
    - \tr \left[\bfW^T\bfzeta^T\bfY \right]
    \right) \right\} \\
    &= \exp \left\{ -\frac{1}{2\sigma^2}\left( \tr \left[\bfW\bfW^T\bfzeta^T\bfzeta \right]
    - 2 \tr \left[\bfW\bfY^T\bfzeta \right]
    \right) \right\} \\
    & \propto \exp \left\{ -\frac{1}{2\sigma^2} \tr \left[\bfW\bfW^T \left(\bfzeta - \bfY\bfW[\bfW\bfW^T]^{-1} \right)^T \left(\bfzeta - \bfY\bfW[\bfW\bfW^T]^{-1} \right) \right]
     \right\}
\end{align*}
and
\begin{align*}
    p(\bfxi \mid \bfzeta) &\propto \exp \left\{ -\frac{1}{2\tau^2} \tr  \left[ \left( \bfzeta - \bfX \bfxi \right)^T \left( \bfzeta - \bfX \bfxi \right) \right] \right\}  \\
    & \propto \exp \left\{ -\frac{1}{2\tau^2}\left( \tr \left[\bfxi^T\bfX^T\bfX\bfxi \right]
    - \tr \left[\bfzeta^T\bfX\bfxi \right]
    - \tr \left[\bfxi^T\bfX^T\bfzeta \right]
    \right) \right\}  \\
    & \propto \exp \left\{ -\frac{1}{2\tau^2} \tr \left[ \left(\bfxi - [\bfX^T\bfX^T]^{-1}\bfX^T\bfzeta \right)^T \bfX^T\bfX \left(\bfxi - [\bfX^T\bfX^T]^{-1}\bfX^T\bfzeta] \right) \right]
     \right\} ,
\end{align*}
leading to 
$$
\bfzeta \mid \bfY \sim \mvn_{N,L} \left( \bfY \bfW^T [ \bfW \bfW^T ]^{-1}, \mathbf{I}_{N}, \sigma^2 [ \bfW \bfW^T ]^{-1} \right)
$$
and 
$$
\bfxi \mid \bfzeta \sim \mvn_{J,L} \left( [\bfX^T \bfX]^{-1}\bfX^T\bfzeta ,  [\bfX^T\bfX]^{-1}, \tau^2 \mathbf{I}_{L} \right).
$$
Since $\E\left\{ \bfzeta \mid \bfY \right\} = \bfY\bfW[\bfW\bfW^T]^{-1} = \hat{\bfzeta}$, we have
\begin{align*}
p_t(\bfxi \mid \bfY) = \mvn_{K,L} \left(  \tilde{\bfxi}(\bfY) ,  [\bfX^T\bfX]^{-1}, \tau^2 \mathbf{I}_{L} \right) .
\end{align*}
where $\tilde{\bfxi}(\bfY) = [\bfX^T \bfX]^{-1}\bfX^T\bfY \bfW^T [ \bfW \bfW^T ]^{-1}$.  Like $\hat{\bfxi}(\bfY)$, $\tilde{\bfxi}(\bfY)$ is unbiased:
\begin{align*}
    \E \left\{ \tilde{\bfxi}(\bfY) \right\} &= [\bfX^T\bfX]^{-1}\bfX^T \E \left\{\bfY \right\} \bfW^T [\bfW \bfW^T]^{-1} \\
    &= [\bfX^T\bfX]^{-1}\bfX^T \bfX \bfxi\bfW  \bfW^T [\bfW \bfW^T]^{-1} = \bfxi . 
\end{align*}

More generally, 
\begin{align*}
    \tilde{\bfxi}(\bfY)  &\sim \mvn \left(  \bfxi, [\bfX^T\bfX]^{-1}, [\bfW\bfW^T]^{-1}
    \bfW
    \bfOmega_{*}
    \bfW^T 
    [\bfW\bfW^T]^{-1}
    \right)
\end{align*}
where $\bfOmega_{*} = \tau^{2}_{*} \bfW^T\bfW + \sigma^{2}_{*} \mathbf{I}_{J}$ and $ \tau^{2}_{*}$ and  $\sigma^{2}_{*}$ are the  values of $\tau^2$ and $\sigma^2$ under the true data generation mechanism.

\subsection{Cut posterior}\label{ap:derivationcutpost}

The cut posterior is defined as:
\begin{align*}
    p_c(\bfxi \mid \bfY) = \int p(\bfxi \mid \bfzeta) p(\bfzeta \mid \bfY) \dd \bfzeta
\end{align*}
We can use a similar trick to the one is Section \ref{ap:derivationfullpost}.  In particular note that, 
\begin{align*}
    \bfxi \mid \bfY &= [\bfX^T\bfX]^{-1} \bfX^{T} \left\{  \bfY \bfW^{T} \left[ \bfW \bfW^T \right]^{-1} + \bfEta^{*}\right\} + \bfEpsilon^{*} \\
    &= [\bfX^T\bfX]^{-1} \bfX^{T}\bfY \bfW^{T} \left[ \bfW \bfW^T \right]^{-1} + [\bfX^T\bfX]^{-1} \bfX^{T} \bfEta^{*} + \bfEpsilon^{*}
\end{align*}
where $\bfEta^{*} \sim \mvn_{I,L} \left( \mathbf{0}, \mathbf{I}_N, \sigma^2 [\bfW\bfW^T]^{-1}\right)$ and $\bfEpsilon^{*} \sim \mvn_{K,L} \left( \mathbf{0}, [\bfX^T \bfX]^{-1}, \tau^2 \mathbf{I}_L\right)$.

Now, using again the properties discussed at the beginning of this section, we have $[\bfX^T\bfX]^{-1} \bfX^{T} \bfEta^{*} \sim \mvn_{K,L} \left(\mathbf{0}, [\bfX^T\bfX]^{-1}, \sigma^2 [\bfW\bfW^T]^{-1}\right)$, and  $[\bfX^T\bfX]^{-1} \bfX^{T} \bfEta^{*} + \bfEpsilon^{*} \sim \mvn \left( \mathbf{0}, [\bfX^T\bfX]^{-1}, \sigma^2 [\bfW\bfW^T]^{-1} + \tau^2 \mathbf{I}_L\right)$, leading to
\begin{align*}
p_c\left(\bfxi \mid \bfY\right) = \mvn_{K,L} \left( \tilde{\bfxi}(\bfY)  , [\bfX^T \bfX]^{-1} , \sigma^2 [\bfW \bfW^T]^{-1} + \tau^2 \mathbf{I}_{L} \right) .
\end{align*}
